\documentclass{aa}  

\usepackage{graphicx}
\usepackage{txfonts}
\usepackage{hyperref}

\usepackage{ulem}
\usepackage{placeins}
\usepackage[dvipsnames]{xcolor} 
\usepackage{float}

\usepackage{natbib}
\bibpunct{(}{)}{;}{a}{}{,} % to follow the A&A style

\usepackage{supertabular}
\usepackage{booktabs}
\usepackage{longtable}

\begin{document}

\title{A blue-straggler merger scenario origin for the 
\texorpdfstring{$\gamma$}{gamma}~Persei binary system}

   \author{Dóra Tarczay-Nehéz
          \inst{1,2}\thanks{\email{tarczaynehez.dora@csfk.org}}
          \and
          László Molnár
          \inst{1,2,3}
          \and
          Rozália Z.~Ádám\inst{1,2,3}}

   \institute{Konkoly Observatory, HUN-REN CSFK, MTA Centre of Excellence, Budapest, Konkoly Thege Mikl\'os \'ut 15-17, Hungary
         \and
             MTA–HUN-REN CSFK Lendület "Momentum" Stellar Pulsation Research Group
        \and
            ELTE E\"otv\"os Lor\'and University, Institute of Physics and Astronomy, 1117, P\'azm\'any P\'eter s\'et\'any 1/A, Budapest, Hungary
             }

   \date{Received \today{}; accepted XXX, 2026}

 \abstract{We used \texttt{MIST} isochrone fitting and a dedicated grid of stellar evolution models computed with \texttt{MESA} to constrain the ages of the components of the $\gamma$~Persei binary system. 
While individual stars can be matched to the models at specific metallicities, no joint isochrone solution reproduces both the observed masses and evolutionary states.

The stellar evolutionary tracks calculated by \texttt{MESA} reveal a clear evolutionary mismatch.
The primary component of the system is in a post-main-sequence phase consistent with the red giant branch or red clump.
In contrast, the lighter secondary component lies near the turn-off point of the main sequence or is in the early phase of the subgiant branch.

This discrepancy can be overcome by assuming that the $\gamma$ Persei system was born as a triple and the primary component is a rejuvenated star formed through a merger of a close-by pair of main-sequence stars.
We show that the merger must have occurred no later than a few hundred megaryears after system formation, and the progenitor masses of the merging stars are restricted by a combination of stars that fall within a narrow band in the $(M_{1,a},M_{1,b})$ plane, corresponding to $M_{1,a}\simeq0.9$--$2.1\,M_\odot$ and $M_{1,b}\simeq2.3$--$2.5\,M_\odot$.}

\keywords{stars: evolution -- stars: binaries: general -- stars: individual: $\gamma$~Persei -- stars: mergers -- stars: blue stragglers -- Hertzsprung--Russell diagram, methods: numerical}

   \maketitle
%
%-------------------------------------------------------------------

\section{Introduction}

$\gamma$ Persei is a known, long-periodic spectroscopic binary system with an orbital cycle of $\sim14.6$ years.
Its binary nature was first mentioned by \cite{Mauryetal1897} and \cite{Campbell1908,Campbelletal1909} from spectroscopy, and the orbital period was first determined by \cite{McLaughlin1948} from radial velocity measurements, who also suggested that the system should be close to $90^\circ$ inclination. 
Yet, although the system is bright enough to be visible to the naked eye, the first total eclipse was not observed until 1990 by \cite{Griffinetal1994}.
The system was first resolved interferometrically by \citet{Wilson-1941}, and later by \citet{Labeyrie-1974} and \citet{McAlister-1982}, and a visual orbital solution indicating a nearly edge-on orbit was calculated by \citet{Popper-1987}. 
The masses of the components were initially determined to be $M_1 = 4.73$\,M$_\odot$ and $M_1 = 2.75$\,M$_\odot$, both by \citet{McLaughlin1948} and then by \citet{McAlister-1982}, but then \citet{Popper-1987} derived significantly lower values of $M_1 = 3.0$\,M$_\odot$ and $M_1 = 2.0$\,M$_\odot$.

The system has since been the focus of diverse research using several observational methods, which include both visual \citep{Hartkopf2001} and spectroscopic observations \citep{McWilliam1990,Pourbaixetal1999,Pourbaixetal2000,Pourbaixetal2004,Piccottietal2020,Diamantetal2023}, revealing a slight metallicity deficit, compared to the Sun ([Fe/H] $\simeq -0.19 - (-0.20)$ in the literature).
Moreover, \cite{Pourbaixetal2004} classified both components as luminosity class III giants based on detailed spectroscopic data.
New orbital elements of the system were calculated by \cite{Pourbaixetal1999}, confirming that the system has a highly eccentric orbit ($e=0.785$). 
Based on its parallax ($14.1252"$), the distance of $\gamma$ Persei is measured to be $\sim70.8$ pc \citep{GaiaEDR3}.
\citet{Diamantetal2023}, which was detected the chromosphere of the giant component.

The system is a member of the $\zeta$ Aurigae class \citep{AkeandGriffin2015}.
This class consists of double-lined binaries with a cool G–K giant primary and a hotter A-type secondary.
In the case of $\gamma$ Persei, the primary component is a luminous G-type giant with a Johnson V-band magnitude of $2.91$ \citep{Ginestet2002}, dominating the system’s brightness. 
The secondary star is significantly fainter, with an estimated V magnitude of $4.54$.
The system’s total brightness is 
$V=2.93$ and $B=3.63$ mag, with eclipse depths of $0.28$ mag in V and $0.54$ mag in B \citep{Griffinetal1994}.
These values yield to the individual color indices of $(B-V )= +0.96$ mag for the primary and $+0.11$ mag for the companion component, which is consistent with a G-type and an A-type classification, respectively. 

Following the first documented 1990 primary eclipse, the next event in 2019 was reported by \cite{Diamantetal2023}, enabling a precise refinement of the orbital period. 
The previous eclipse in 2005 occurred close to the solar conjunction, and thus it was impossible to observe from Earth.
It was later recovered and published in \cite{AdammandMolnar2025} based on archival data from the Solar Mass Ejection Imager (SMEI), thereby filling the observational gap between the two known events and further strengthening the orbital period determination.

The 2019 eclipse was observed using DSLR CMOS photometry, which was calibrated to the 1990 Johnson B-band light curve using spectrophotometric reference data \citep{Diamantetal2023}. 
This approach allowed the DSLR measurements to be transformed onto a standard photometric scale, enabling direct comparison with the earlier eclipse and precise timing analysis.
The calibration was sufficient to refine the orbital period to $5329.0\pm0.05$ days, which is in agreement with the previous spectroscopic data. 
Additional partial eclipse phases were sampled by Telescopio Internacional de Guanajuato Robótico-Espectroscópico (TIGRE) spectra and Transiting Exoplanet Survey Satellite (TESS) photometry, the latter resolving the first contact phase and totality with high cadence, ending right around fourth contact. 
From the eclipse geometry and projected orbital velocity ($\sim40$ km/s), the radius of the primary star was derived to be $22.7\,R_\odot$, while the radius of the secondary component was calculated to be $3.9\,R_\odot$. TESS observed the star later again, most notably for two consecutive sectors in late 2024, making a global asteroseismic analysis possible. A seismic mass of $3.27\pm0.13\,M_\odot$ was determined for the primary component based on the detected solar-like oscillations, which we present in a companion paper \citep{Adametal2026}.

While $\gamma$ Persei is a well-characterized binary system and is the focus of extensive investigations, studies have drawn attention to a puzzling evolutionary mismatch between its components. 
The spectral types determined for the components by \citet{Bahng-1958} and \citet{Cowley-1976} indicated a G--K-type red-giant primary and an A--type secondary.
This contradicts the expectations for a coeval binary system. \citet{McAlister-1982} already alluded to this and tried to resolve the issue by suggesting that the secondary should be classified as a hotter, B--type star. 
Later \citet{Pourbaixetal1999} and \cite{Griffin2007} both pointed out again that the less massive A-type companion appears to be more evolved than the main G-type component. 
To overcome this discrepancy, several competing interpretations have been raised, including mass loss, dynamical capture, the primary being a close binary \citep{Popper-1987}, early-stage mergers, and episodic mass transfer during the giant’s tip-RGB phase \citep[see][]{Diamantetal2023}. 

In this study, we examined whether a formation pathway exists in which both stars could have formed simultaneously and evolved jointly without significant disturbance, consistent with the present-day configuration of $\gamma$ Persei.
In the following, we describe the applied methods used to map the parameter space of potential co-evolutionary configurations, focusing on metallicity ($Z$), initial mass ($M$), mixing-length parameter ($\alpha_\mathrm{MLT}$), convective overshoot, and rotational velocity ($v_\mathrm{rot}$), using both \texttt{MIST} and \texttt{MESA} evolutionary models (Section~\ref{sec:methods}). 
Section~\ref{sec:results} presents our results, Section~\ref{sec:discussion} provides a discussion on the results, and the paper closes with a conclusion for the system’s formation history in Section~\ref{sec:conclusion}.

%--------------------------------------------------------------------

\begin{table}
    \renewcommand{\arraystretch}{1.2}
    \centering
    \caption{Physical parameters of the $\gamma$ Persei ($\gamma$ Per) binary system.}
    \label{tab:gammapersei_params}
    \begin{tabular}{lcc}
        \hline
        \textbf{Parameter} & \textbf{Primary} & \textbf{Secondary} \\[1ex]
        \hline
        \\
        $T_\mathrm{eff}^1$ & $4970\pm70$ K & $8400 \pm70$ K \\
        $(\log L/L_\odot)^1$ & $2.45 \pm0.06$ & $1.83 \pm0.06$ \\
         $M$ [$M_{\odot}$]$^{1,2,3,4}$ & $3.45\pm 0.35$ & $2.4\pm0.2$ \\
        $R$ [$R_\odot$]$^1$ & $22.7\pm1.14$ & $3.9\pm0.2$ \\
        
        [Fe/H]$^{3,5}$ & $-0.19\pm0.08$ dex & $-0.2\pm0.08$ dex\\[2ex]
        Period [days]$^1$ & \multicolumn{2}{c}{$5329.0\pm0.05$}  \\[1ex]
        \hline
        \\
        \multicolumn{3}{l}{
            \parbox[t]{0.95\linewidth}{
                \footnotemark[1]\cite{Diamantetal2023}; \footnotemark[2]\cite{Griffinetal1994};\\
                \footnotemark[3]\cite{AdammandMolnar2025};
                \footnotemark[4]\cite{Adametal2026};\\
                \footnotemark[5]\cite{RosasPortilla2024}.
            }
        }\\     
    \end{tabular}
\end{table}

\begin{table*}[ht]
    \renewcommand{\arraystretch}{1.2}
    \centering
    \caption{Summary of the input parameters of \texttt{MESA} simulations.}
    \label{tab:mesa_parameters}
    \begin{tabular}{llll}
    \hline
    \textbf{Parameter} & \textbf{Value / Range} & \textbf{Notes} & \textbf{Count} \\
    \hline
    \multicolumn{4}{l}{\textbf{Stellar composition and the initial parameters for the primary component}} \\[0.5ex]
    \hline
    Initial mass ($M_*$) & $3.1-3.8\,M_\odot$ & $\Delta M = 0.1\,M_\odot$  & 8 values\\
    Initial metallicity ([Fe/H]) & $-0.29$, $-0.19$, $-0.09$, --0.036 && 4 values \\
    Initial helium fraction ($Y_{\text{init}}$) & $0.256$ & Fixed \\
    Initial hydrogen fraction ($X_{\text{init}}$) & $1 - Z - Y$ & Computed dynamically \\
    Stopping condition & TACHeB phase & \\
    
    \hline
    \multicolumn{4}{l}{\textbf{Stellar composition and the initial parameters for the secondary component}} \\[0.5ex]
    \hline
    Initial mass ($M_*$) & $2.2 - 2.6 M_\odot$ & $\Delta M = 0.1\,M_\odot$  & 5 values\\
    Initial metallicity ([Fe/H]) & $-0.30$, $-0.20$, $-0.10$, --0.045 && 4 values \\
    Initial helium fraction ($Y_{\text{init}}$) & $0.256$ & Fixed & \\
    Initial hydrogen fraction ($X_{\text{init}}$) & $1 - Z - Y$ & Computed dynamically \\
    Stopping condition & tip of the Red Giant Branch & \\
    \hline
    \multicolumn{4}{l}{\textbf{Mixing and convection physics}} \\[0.5ex]
    \hline
    Mixing length parameter ($\alpha_{\text{MLT}}$) & $1.5-2.5$ & $\Delta \alpha_\mathrm{MLT} = 0.2$ &  6 values\\
    Semi-convection parameter ($\alpha_s$) & $0.1$ &  & \\
    Convective boundary criterion & Ledoux & \\
    Convective overshoot scheme & Exponential & \\
    Convective overshoot parameters ($f$, $f_0$) & See Table\,\ref{tab:overshoot} & Varies per set & 3 values \\
    \hline
    \multicolumn{4}{l}{\textbf{Opacity sources and nuclear network}} \\[0.5ex]
    \hline
    High-$T$ opacity source & OPAL & \\
    Low-$T$ opacity source & \citet{Fergusonetal2005} &  opacity prefix (\texttt{lowT\_fa05})  \\
    Heavy element mixture & \citet{Asplundetal2009} & Solar-like abundance (`A09') \\
    Carbon/Oxygen opacities & Enabled &  \\
    Type2 opacities & Enabled &  \\
    Nuclear network & \texttt{pp\_and\_cno\_extras\_o18\_ne22} & Tracks 19 isotopes \\
    
    \hline
        \textbf{Total Models} & & &936 \\
        \hline
    \end{tabular}
\end{table*}

\begin{table}[ht]
    \renewcommand{\arraystretch}{1.2}
    \centering
    \caption{Overshoot coefficients ($f$, $f_0$) adopted in this work for convective core and neighboring radiative zones.}
    \label{tab:overshoot}
    \begin{tabular}{lcc}
        \hline
        \textbf{Region} & \textbf{$f$} & \textbf{$f_0$} \\
        \hline
        \multicolumn{3}{c}{\textbf{Set A — Low Overshoot}} \\
        \hline
        Core (H-burning) & 0.012 & 0.002 \\
        Radiative shell  & 0.022 & 0.002 \\[1ex]
        \hline
        \multicolumn{3}{c}{\textbf{Set B — Moderate Overshoot}} \\
        \hline
        Core (H-burning) & 0.020 & 0.004 \\
        Radiative shell  & 0.030 & 0.004 \\[1ex]
        \hline
        \multicolumn{3}{c}{\textbf{Set C — High Overshoot}} \\
        \hline
        Core (H-burning) & 0.030 & 0.005 \\
        Radiative shell  & 0.040 & 0.005 \\
        \hline
    \end{tabular}
\end{table}

\section{Methods} \label{sec:methods}
 
\subsection{\texttt{MIST-}based isochrone model setup}
  
To validate the coevolution of the binary system of the $\gamma$ Persei, the  2.1 release of the isochrone package of the \texttt{MESA} Isochrones \& Stellar Tracks (\texttt{MIST}) database \citep{MIST0, MIST1} was utilized.
The package offers a comprehensive grid for stellar evolutionary modelling, providing isochrones and synthetic photometry across a broad range of physical parameters. 
The full metallicity ([Fe/H]) range supported by \texttt{MIST} spans from $-4.0$ to $+0.5$ dex. 
In our analysis, we varied metallicity with [Fe/H] values from $-0.2$ and $-0.19$ down to $-2.0$ dex to probe lower metallicity regimes as well. 

Synthetic magnitudes were extracted for the \textit{Gaia} photometric system, with extinction ($A_\mathrm{v}$) fixed at $0.0$ for all models, although the database covers a range of $A_\mathrm{v}$ between $0$ and $6$. 
Isochrones for both rotational configurations available in the grid were investigated: $v/v_\mathrm{(crit)} = 0.0$, representing nonrotating stars, and $v/v_\mathrm{(crit)} = 0.4$, corresponding to moderately rotating stars. 
Here, $v_\mathrm{(crit)}$ denotes the critical rotation velocity at which centrifugal acceleration at the stellar equator balances gravitational acceleration. 
Exceeding this threshold, the star would be disrupted by centrifugal forces. 

For the zeroth-order constraint, we adopted the standard \texttt{MIST} age grid, which consists of 107 discrete age points uniformly spaced in $\log({\rm Age}\, [yr])$ from $5.0$ to $10.3$ dex in $0.05$ dex increments. 
Upon identifying the subset of isochrones that provided satisfactory fits to both stellar components, we refined the age resolution locally by resampling the grid in steps of $0.01$ dex, allowing for a more precise age determination of the system.

\subsection{\texttt{MESA} stellar evolution tracks}

\label{sec:mesa_setup}
In addition to the parameter space offered by the \texttt{MIST} grid (e.g., rotation and metallicity), we performed a set of stellar evolution calculations using the stable version of 23.05.1 of the open-source Modules for Experiments in Stellar Astrophysics (\texttt{MESA}) software package \citep{Paxton2011,Paxton2013,Paxton2015,Paxton2018,Paxton2019,Jermyn2023}.
All evolutionary tracks are available under Zenodo\footnote{\url{https://doi.org/10.5281/zenodo.19470956}} \citep{tarczay_nehez_2026_19470956}.

\subsubsection{Metallicity, heavy-element abundance, and stellar mass grid}
For each component of the binary system, we constructed a mass grid centered at $3.45\pm 0.35\,M_\odot$ and 
$2.4 \pm 0.2\,M_\odot$, respectively, with a resolution of $0.1\,M_\odot$  \citep{ Griffinetal1994, Diamantetal2023, AdammandMolnar2025,Adametal2026}.
This resulted in eight and five mass points per star. 
The primary component was evolved up to the terminal-age core helium-burning (\texttt{TACHeB}, \texttt{stop\_at\_phase\_TACHeB = .true.}) phase, while the secondary was evolved to the tip of the red giant branch (RGB, \texttt{stop\_at\_phase\_He\_Burn = .true.}).
We adopted a solar composition for both stars and varied the metallicity separately for each star: [Fe/H] = $-0.19$ for the primary and [Fe/H] = $-0.20$ for the secondary, each with a local grid of $\pm0.1$ dex in $0.1$ dex steps \citep{RosasPortilla2024, AdammandMolnar2025}. 
We note, however, that the [Fe/H] values derived by \citet{McWilliam1990} were based on the abundance tables of \citet{Grevesse-1984}, which provided a solar metallicity of $Z=0.02$. 
Assuming the relation of
\begin{equation}
    Z = Z_\odot\cdot 10^{[\rm Fe/H]},    
\end{equation}
the adopted metallicity values correspond to $Z\simeq 0.00126$ and $Z\simeq 0.00129$, respectively.
Consequently, when recalculated using the MESA default solar metallicity of $Z_\odot=0.014$ \citep{Asplundetal2009}, these translate to revised [Fe/H] values of $-0.045$ and $-0.036$, respectively. 
Because this recalibration leads to metallicity values that differ from our originally adopted grid, we computed an additional set of evolutionary models at $[\mathrm{Fe/H}]=-0.045$ and $-0.036$ to ensure consistency with the MESA abundance scale.
Altogether, this yields four metallicity values for each component.
The parameter \texttt{Z\_{base}} was dynamically maintained equal to the initial metallicity, \texttt{Z\_{init}}.

\subsubsection{Opacity tables}
In this work, the stellar energy transport efficiency was characterized using the built-in \texttt{MESA} opacity tables. 
For high-temperature conditions, the Opacity Project at Livermore (OPAL) data of \cite{IglesiasandRogers1993, Iglesiasetal1996} were employed, specified by the \texttt{kap\_file\_prefix = "a09"} parameter. 
Conversely, the low-temperature regime was parameterized by the inclusion of opacity data from \cite{Fergusonetal2005} by using \texttt{kap\_lowT\_prefix = "lowT\_fa05\_a09p"}.
We used specialized opacity tables—namely those enriched in carbon and oxygen (\texttt{kap\_CO\_prefix = "a09\_co"}) and the type 2 opacities (\texttt{use\_Type2\_opacities = .true.})
It should be emphasized, however, that their overall influence on the evolutionary trajectories of the stars is negligible. 

\subsubsection{Convection-core overshoot and the mixing length parameter}
Convection was modeled across all computations using the mixing length theory (MLT) coupled with the time-dependent convection (\texttt{MLT\_option = "TDC"}). Following the approach of, for example, \citet{XuandLi2004a,XuandLi2004b} and \citet{JoyceandTayar2023}, the Ledoux criterion (\texttt{use\_Ledoux\_criterion = .true.}) was employed to define convective boundaries, utilizing a semi-convection parameter ($\alpha_{s}$) of $0.1$ (\texttt{alpha\_semiconvection = 0.1}).

We systematically investigated the MLT parameter ($\alpha_\mathrm{MLT}$), varying its value from $1.5$ to $2.5$ with a resolution of $0.2$ (six distinct values). 
For each $\alpha_\mathrm{MLT}$ setting, three distinct overshoot prescriptions were tested using an exponential overshoot scheme (\texttt{overshoot\_scheme = "exponential"}). 
This scheme was applied below the hydrogen-burning shell (\texttt{overshoot\_zone\_type(1) = "burn\_H"}) and above the "non-burning" regions (\texttt{overshoot\_zone\_type(2) = "none"}). 
Following the most recent works of \cite{Ziolkowska2024} and \cite{TND2026}, we used the three distinct overshoot prescriptions for all models. 
The applied input parameters are summarized in Tables\,\ref{tab:mesa_parameters} and \ref{tab:overshoot}.

\subsubsection{Nuclear-reaction network}
For the analysis of intermediate-mass stars with initial masses between $2$ and $4\,M_{\odot}$, we adopted the extended nuclear network \texttt{pp\_and\_cno\_extras\_o18\_ne22}. 
Reaction rates in this network were taken from the Nuclear Astrophysics Compilation of Reaction Rates \cite[NACRE; ][]{Angulo1999} and the Joint Institute for Nuclear Astrophysics REACtion LIBary (JINA REACLIB) database \citep{Cyburt2010}.
The network follows 19 isotopes involved in the proton–proton chain, the carbon–nitrogen–oxygen cycles (including hot CNO branches), and additional burning channels involving heavier nuclei. 
The tracked isotopes are $^1$H, $^2$H, $^3$He, $^4$He, $^7$Li, $^7$Be, $^8$B, $^{12}$C, $^{14}$N, $^{14}$O, $^{16}$O, $^{18}$O, $^{19}$F, $^{18}$Ne, $^{19}$Ne, $^{20}$Ne, $^{22}$Ne, $^{22}$Mg, and $^{24}$Mg.

\subsubsection{Spatial and temporal resolution}
The spatial and temporal resolution of the models can be adjusted by modifying \texttt{MESA}'s default refinement parameters, namely \texttt{mesh\_delta\_coeff} and \texttt{time\_delta\_coeff} \citep[see, e.g.,][]{LiandJoyce2025}. 
These coefficients scale the internal adaptive mesh refinement (AMR) scheme used by \texttt{MESA}. 
In the present work, we retained the default resolution settings, with \texttt{mesh\_delta\_coeff = 1.0} and \texttt{time\_delta\_coeff = 1.0}.

For improved numerical stability and energy conservation, the stellar energy equation was solved using the \texttt{dedt} formulation (\texttt{energy\_eqn\_option = "dedt"}). 
In addition, the \texttt{make\_gradr\_sticky\_in\_solver\_iters} option was activated to enhance convergence at the interfaces between convective and radiative regions \citep[see ][]{Paxton2018,Paxton2019}.

\subsection{Matching ages for different evolutionary tracks}
\label{sec:match_mesa_ages}

In order to identify which combinations of model configurations yield a consistent age for both stellar components, we used a dedicated Python \citep{Python3} script \citep[based on pandas, numpy, and matplotlib;][]{pandas,numpy,matplotlib}, to compare evolutionary tracks and extract age-matching solutions.

\begin{table*}
    \renewcommand{\arraystretch}{1.2}
    \centering
    \caption{Age and mass ranges where the two components of the system intersect \texttt{MIST} isochrones (within the observational errors), grouped by metallicity. Ages are given in megayears. }
    \begin{tabular}{c c c c c}
    \hline
    [Fe/H] & Primary: $t$ ranges [Myr] & Primary: $M$ ranges [$M_\odot$] & Secondary: $t$ ranges [Myr] & Secondary: $M$ ranges [$M_\odot$] \\
    \hline
    $-4.00$ & 8\,900--20\,000 & 0.696--0.882 & 1\,410--1\,580,\ 5\,010 & 1.471--1.527,\ 1.025 \\
    $-3.50$ & 7\,080--20\,000 & 0.699--0.943 & 1\,260--1\,410,\ 4\,470--5\,010 & 1.524--1.580,\ 1.021--1.062 \\
    $-3.00$ & -- & -- & -- & -- \\
    $-2.50$ & -- & -- & -- & -- \\
    $-2.00$ & 1\,000--20\,000 & 0.652--1.707 & 794--891 & 1.745--1.812 \\
    $-1.50$ & 724--2\,820 & 1.235--1.958 & 708--871 & 1.799--1.932 \\
    $-1.00$ & 282--794 & 2.080--2.909 & 708--794 & 1.894--1.983 \\
    \textbf{$-$0.20} & 178--355 & 3.165--3.904 & 631--759 & 2.170--2.338 \\
    \textbf{$-$0.19} & 178--355 & 3.171--3.909 & 631--708 & 2.247--2.343 \\
    \textbf{$-$0.10} & -- & -- & 631--759 & 2.216--2.370 \\
    \textbf{$+$0.00} & 158--282 & 3.501--4.148 &  562--708 & 2.322--2.484\\
    \hline
    \end{tabular}
\label{tab:mist_results}  
\end{table*}
    
\vskip 1em
\noindent The analysis consists of the following steps:
\begin{itemize}
    \item First, it identifies where each stellar component intersects the calculated evolutionary track beyond the pre-main-sequence phase within the observational uncertainties listed in Table\,\ref{tab:gammapersei_params}. 
    \item In cases where no track point falls explicitly within the observational error box, but the track segment crosses it, the script estimates the intersection age via linear interpolation between the segment endpoints. This allows us to identify age-matching solutions even when no direct data point lies within the uncertainty region.
    \item Finally, the script compares the resulting age distributions for both stars and identifies model configuration pairs that yield the smallest age difference. These are considered the best candidates for a physically consistent binary solution.
\end{itemize}
As an additional validation step, we applied a threshold of 0.1\,dex in $\log_{10}$(age) to identify model configuration pairs with sufficiently consistent ages. 
This can be calculated as follows:
\begin{equation}
\label{eq:dex_calc}
    \Delta_{\log} = \log_{10}\left(\frac{t_2}{t_1}\right).
\end{equation}
This ensures that the age estimates derived from two stellar evolution models for the components of the $\gamma$~Persei system differ by no more than 0.1\,dex (corresponding to $\sim$25\%), thereby allowing us to retain only configurations in which both stars can be considered coeval within the adopted tolerance.

To quantify how well individual \texttt{MESA} evolutionary models reproduce the observed properties, we define a scalar goodness-of-fit metric: $\varepsilon$.
This measures the distance between the model predictions and the observational constraints in the three-dimensional parameter space of the effective temperature, luminosity, and radius. 
This metric was only computed for model points that were identified as matching with the observations according to the procedure described in Section\,\ref{sec:match_mesa_ages}.
$\varepsilon$ reads as
\begin{equation} \label{eq:epsilon_def} 
    \varepsilon = \sqrt{ \left(\frac{\Delta \log T_{\mathrm{eff}}}{\sigma_{\log T}}\right)^2 + \left(\frac{\Delta \log L}{\sigma_{\log L}}\right)^2 + \left(\frac{\Delta R}{\sigma_{R}}\right)^2 }, 
\end{equation} 
where $\Delta \log T_{\mathrm{eff}}$, $\Delta \log L$, and $\Delta R$ are the differences between the calculated model parameters and the observed quantities, and $\sigma_{\log T}$, $\sigma_{\log L}$, and $\sigma_R$ are the corresponding observational uncertainties. 

\section{Results} \label{sec:results}
In this section, we present our results on matching the ages of the two components of the $\gamma$ Persei binary system.

\subsection{MIST isochrone fitting}
As a first approach, we examined the stellar parameters of the two stars in the system using
MIST isochrones. 
The results of this analysis are listed in Table\,\ref{tab:mist_results}.

While the secondary component of the system can be consistently matched to \texttt{MIST} isochrones across the full range of metallicities explored, the primary component cannot be fitted in all cases. 
In particular, for metallicities of $\mathrm{[Fe/H]} = -0.1$, $-2.5$, and $-3.0$ the primary component of the $\gamma$~Persei system cannot be fitted on the given isochrones from the \texttt{MIST} database.

Furthermore, a joint fit for both components is only achievable for metallicities $\mathrm{[Fe/H]} = -1.0$ and $\mathrm{[Fe/H]} = -1.5$.
In the case of $\mathrm{[Fe/H]} = -1.5$, both stars can be placed on a single isochrone with ages in the range of $728 - 871$\,Myr. 
In this case, the inferred masses are $1.96$--$1.84\,M_\odot$ for $\gamma$~Persei~1 and $1.93$--$1.80\,M_\odot$ for $\gamma$~Persei~2. 
Similarly, for $\mathrm{[Fe/H]} = -1.0$, a common isochrone solution is found at ages $708-794$\,Myr, yielding masses of $2.17$--$2.08\,M_\odot$ for the primary and $1.98$--$1.89\,M_\odot$ for the secondary.

Although joint isochrone solutions formally exist at $\mathrm{[Fe/H]} = -1.0$ and $-1.5$, these models require stellar masses and metallicities that are inconsistent with the observed values (Table~\ref{tab:gammapersei_params}). Thus, within the range of metallicities and masses allowed by the observations, the MIST isochrones do not provide a coeval configuration for the two components.

\subsection{\texttt{MESA} evolution track fitting}
As the \texttt{MIST} isochrone fitting did not yield any configuration for the system that is consistent with the observed masses and metallicities, we constructed a grid of stellar models with \texttt{MESA}, adopting the parameters described in Section\,\ref{sec:mesa_setup} (see Table\,\ref{tab:mesa_parameters}).

Table\,\ref{tab:mesa_results} lists the results for this analysis.
In general, independently of the adopted physical parameters, we consistently find that the age of the primary component lies in the range of $\sim200-400$~Myr.
In contrast, the secondary component is significantly older, with solutions spanning $\sim550-890$~Myr.
Considering the model pair that minimizes the age difference between the two components, we find ages of $t_1 = 390$~Myr for the primary and $t_2 = 547$~Myr for the secondary component.
This corresponds to an absolute age difference of $156$~Myr.
Using the definition introduced in Eq.~(\ref{eq:dex_calc}), this yields a logarithmic age difference of
\begin{equation}
    \Delta_{\log} \simeq 0.15~\mathrm{dex},
\end{equation}
which corresponds to a relative age difference of  $\sim40$\%.
Therefore, even the closest age pairing identified in our model grid exceeds the adopted tolerance of 0.1~dex and cannot be considered coeval under our selection criterion.

\begin{figure*}
    \centering
    \includegraphics[width=\textwidth]{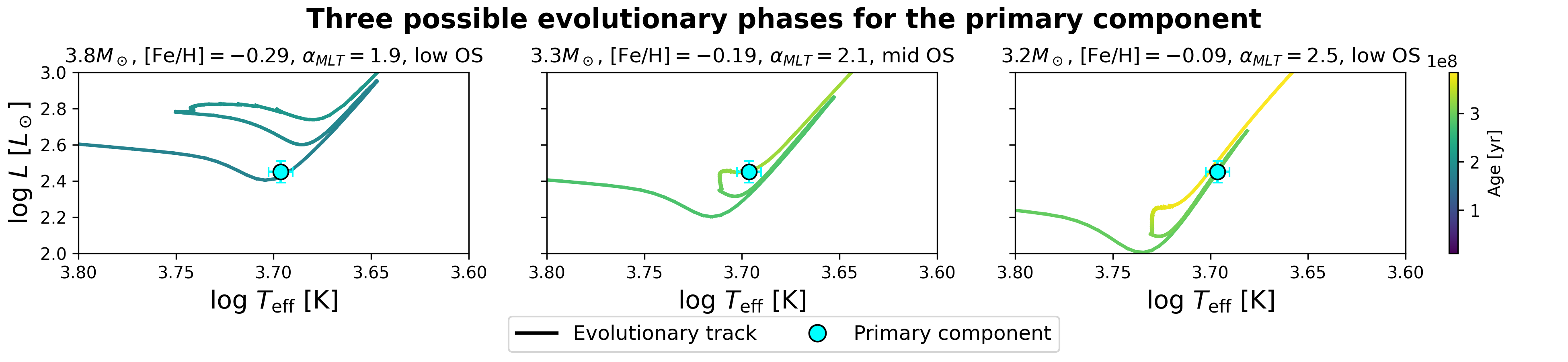}
    \caption{Evolutionary tracks for $\gamma$~Persei primary component, colour-coded by stellar age. The three subpanels illustrate representative models that place the star in distinct post-main-sequence evolutionary phases. From left to right: End of the subgiant branch; the red-clump phase; either the upper red giant branch or the early asymptotic giant branch. The position of the primary component is shown in cyan with observational uncertainties.}
    \label{fig:primary_example_fit}
\end{figure*}

\begin{figure*}
    \centering
    \includegraphics[width=\textwidth]{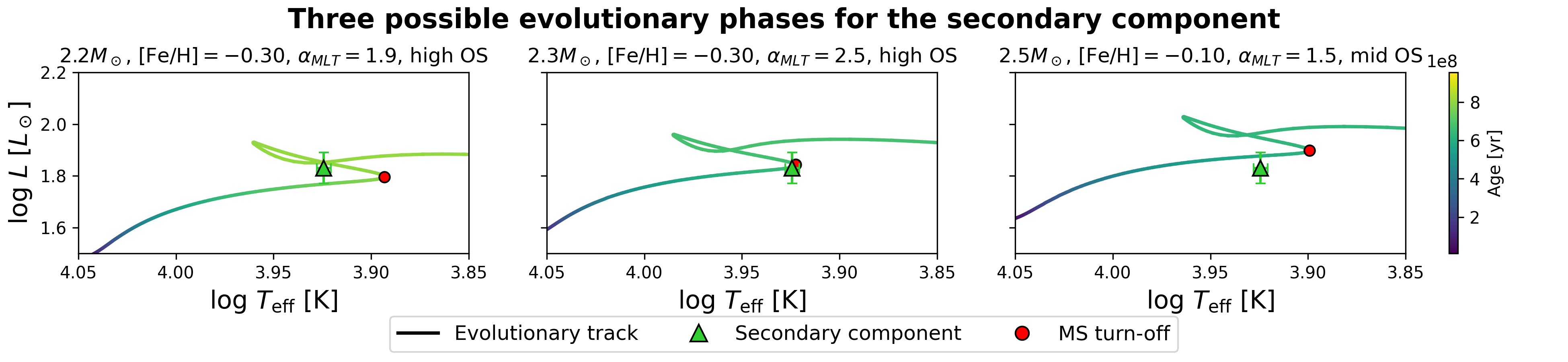}
    \caption{
    Evolutionary tracks for $\gamma$~Persei secondary component, colour-coded by stellar age. 
    The three subpanels illustrate representative models that place the star in distinct evolutionary stages. From left to right: subgiant phase; a model located in the vicinity of the main-sequence turn-off point; a main-sequence phase. 
    The observed position of the secondary component is shown with its measurement uncertainties (green triangle), and the main-sequence turn-off point is marked for each model with a red circle.
    }
    \label{fig:secondary_example_fit}
\end{figure*}

\subsection{Evolutionary state of the two components}

The locations of the two components of the $\gamma$~Persei system were examined in detail using the \texttt{MESA} evolutionary tracks described in Section~\ref{sec:mesa_setup}.
Representative evolutionary tracks are shown in Figures~\ref{fig:primary_example_fit} and~\ref{fig:secondary_example_fit}, with the observationally determined positions of the two stars overplotted.
Appendices \ref{app:primar} and \ref{app:secondary} contain the full set of evolutionary tracks for the whole explored parameter-space (see Figures\,\ref{fig:1app1} -- \ref{fig:1app3} and \ref{fig:2app1} -- \ref{fig:2app3}).

For the primary, we consistently find that its position in the Hertzsprung--Russell diagram (HRD) is near or beyond the red giant branch.
Depending on the adopted model parameters, the star is placed onto the HRD 
(i) near the luminosity dip at the end of the subgiant branch;
(ii) in the red clump phase, corresponding to stable core helium burning; or 
(iii) on the red giant branch or in the early phase of the asymptotic giant branch (see Figure\,\ref{fig:primary_example_fit}).

Their position on the HRD shows only a mild dependence on the adopted input parameters (see Figures \ref{fig:2app1} -- \ref{fig:2app3}).
In general, lower metallicity or larger overshoot values tend to shift the primary component toward the red clump, while higher metallicity or smaller overshoot favor a location near the subgiant-branch dip. 
However, across all combinations of $\alpha_{\rm MLT}$, [Fe/H], and overshoot, the primary remains in a post-main-sequence phase (see Figures\,\ref{fig:1app1}--\ref{fig:1app3}).

Considering the secondary component, the dependence on the input parameters is similarly modest.
However, we found that only models within the $2.2\text{--}2.4\,M_\odot$ range align with the position of the star (see, e.g., Figure\,\ref{fig:2app1}).
Higher mass tracks (e.g.,\ $2.6\,M_\odot$) do not fit the measurements of the star for any combination of $\alpha_{\rm MLT}$, [Fe/H], or overshoot (see Figures\,\ref{fig:2app1}--\ref{fig:2app3}).
This yields a more precise mass determination for this component of the system, effectively tightening the observational limits for the given set of model assumptions.

Nevertheless, all three solutions suggest that the primary component of the system is a post-main-sequence star and must be at least in the red giant branch phase, independently of whether it is currently undergoing core helium burning (red clump) or is located slightly before or after this phase.

In contrast, the secondary component of the $\gamma$ Persei system is in a remarkably different evolutionary stage.
The results of the \texttt{MESA} evolutionary tracks indicate that the position of this component is  either close to the turn-off point of the main sequence or in the very early phase of the subgiant branch (see Figure\,\ref{fig:secondary_example_fit}).

The combination of these evolutionary states implies a substantial discrepancy between the ages inferred for the two components under the assumption of single-star evolution.
While the secondary component requires an age sufficient to reach helium ignition, the primary component remains near the main-sequence turn-off or early subgiant phase.
This evolutionary mismatch persists across the explored range of stellar parameters.

\subsection{The most plausible model configurations based on the \texorpdfstring{$\varepsilon$}{epsilon} metric}
\label{sec:epsilon_results}

We calculated the most plausible models for each component  by utilizing the goodness-of-fit metric, $\varepsilon,$ in the $\log T_\mathrm{eff} - \log L - R$ phase defined in Equation~\ref{eq:epsilon_def}.
For the primary component, the best-fitting models yield $\varepsilon \simeq 0.028$--$0.041$. 
All of these solutions correspond to the red-clump phase (see Figure\,\ref{fig:primary_example_fit}), 
with masses of $M \simeq 3.4$--$3.5\,M_\odot$ and metallicities 
at [Fe/H] $= -0.29$, placing the star in the core helium-burning phase.
The most relevant configurations are summarized in Table~\ref{tab:primary_best_models}.

\begin{figure}
    \centering
    \includegraphics[width=\linewidth]{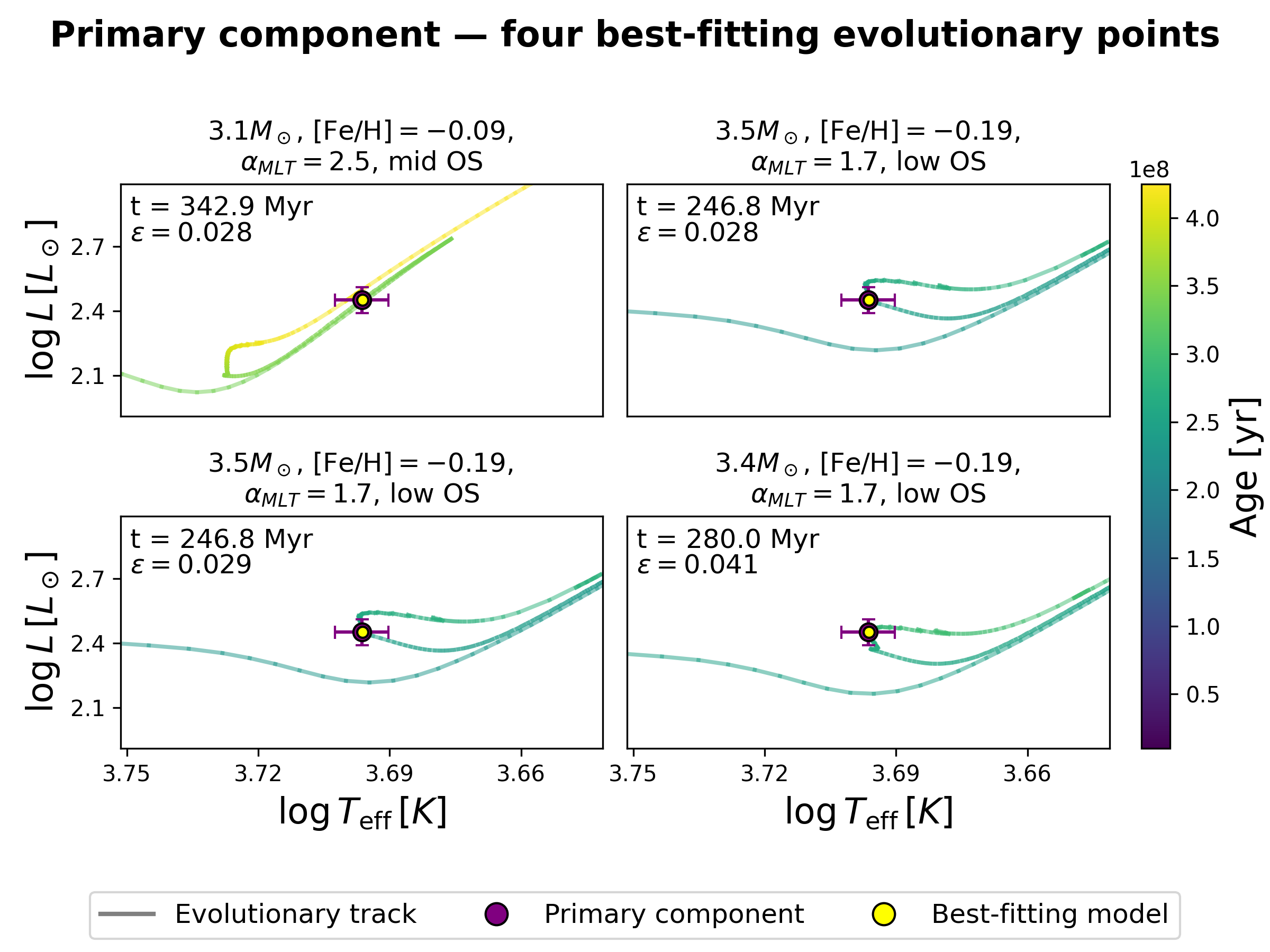}
    \caption{Four best-fitting models for the primary component of the $\gamma$ Persei system. Each model corresponds to the red-clump phase of stellar evolution. The yellow point corresponds to the best-fitting model, while the purple point (with error bars) represents the place of the primary component.}
    \label{fig:best_fit_prim}
\end{figure}

\begin{table}[t]
    \renewcommand{\arraystretch}{1.2}
    \centering
    \caption{Best-fitting \texttt{MESA} models for the primary component, 
    ranked by the goodness-of-fit metric, $\varepsilon$.}
    \label{tab:primary_best_models}
    \begin{tabular}{ccccccc}
        \hline
        $M$ & [Fe/H] & $\alpha_{\mathrm{MLT}}$ 
           & $t$ & $R$ & $\log T_{\mathrm{eff}}$ & $\varepsilon$ \\
           
        [$M_\odot$] & & & [Myr] & [$R_\odot$] & [K] & \\
        \hline
        \hline
        \multicolumn{7}{c}{\textbf{Low overshoot models}} \\
        \hline
        3.1 & $-0.09$ & 2.5 & 343 & 22.679 & 3.6960 & 0.028 \\
        3.5 & $-0.19$ & 1.7 & 247 & 22.672 & 3.6985 & 0.028 \\
        3.5 & $-0.19$ & 1.7 & 248 & 22.680 & 3.6962 & 0.029 \\
        3.4 & $-0.19$ & 1.7 & 280 & 22.703 & 3.6974 & 0.041 \\
        \hline
    \end{tabular}

    \tablefoot{
        All configurations correspond to red-clump (core He-burning) models.
    }
\end{table}

\begin{figure}
    \centering
    \includegraphics[width=\linewidth]{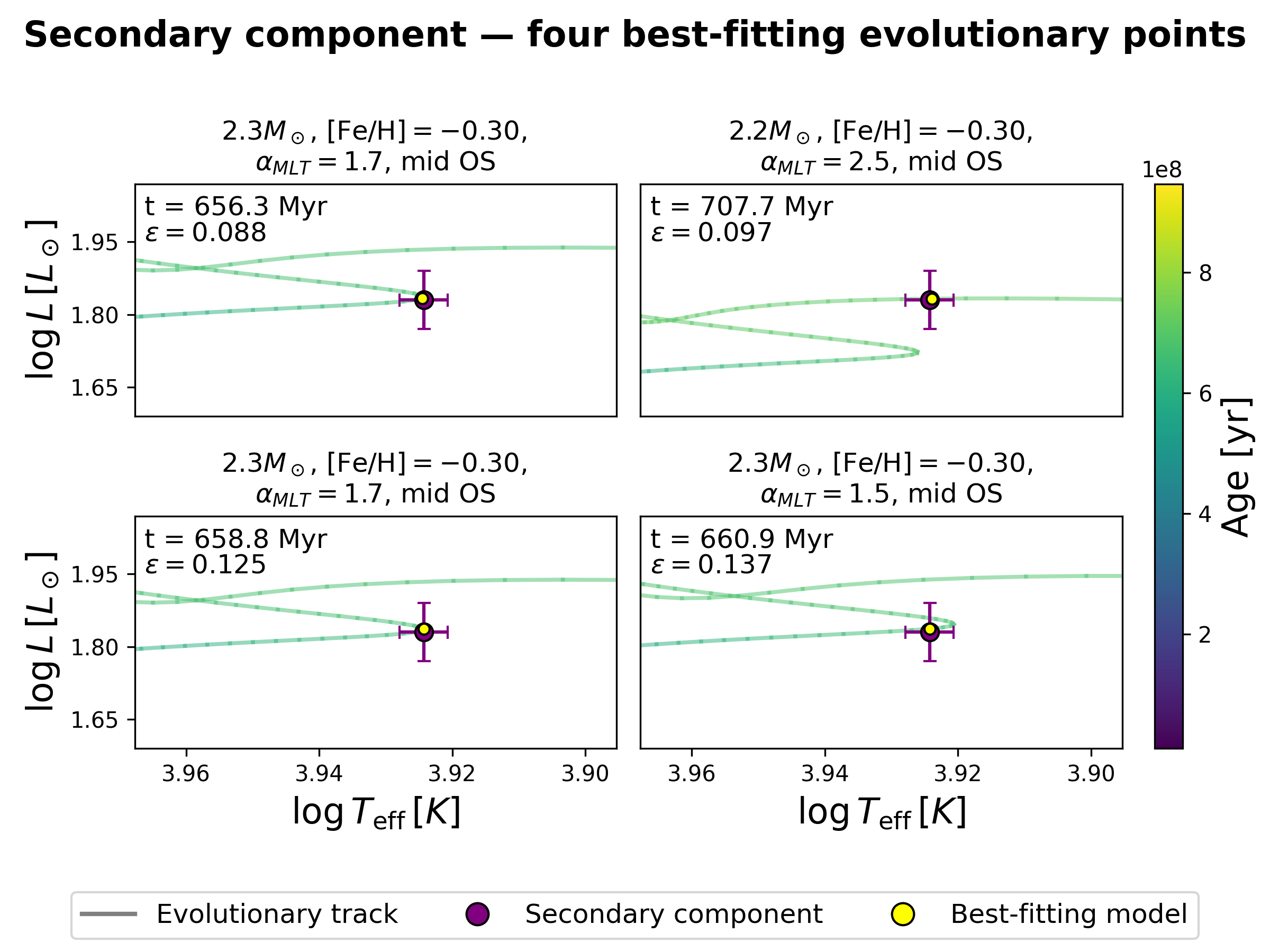}
    \caption{Four best-fitting models for the secondary component of the $\gamma$ Persei system. The panels correspond to different evolutionary stages: the two left panels show models near the main-sequence turn-off (the lower left panel is still on the late main sequence), while the upper right panel represents a subgiant model. The yellow point corresponds to the best-fitting model, while the purple point (with error bars) presents the place of the primary component.}
    \label{fig:best_fit_sec}

\end{figure}

\begin{table}[t]
    \centering
    \renewcommand{\arraystretch}{1.2}
    \caption{Best-fitting \texttt{MESA} models for the secondary component,
    ranked by the goodness-of-fit metric $\varepsilon$. }
    \label{tab:secondary_best_models}
    \begin{tabular}{ccccccc}
        \hline
        $M$ & [Fe/H] & $\alpha_{\mathrm{MLT}}$
           & $t$ & $R$ & $\log T_{\mathrm{eff}}$ & $\varepsilon$ \\
           
        [$M_\odot$] & & & [Myr] & [$R_\odot$] & [K] & \\
        \hline
        \hline
        \multicolumn{7}{c}{\textbf{Medium overshoot models}} \\
        \hline
        2.3 & $-0.30$ & 1.7 & 656.3 & 3.893 & 3.9245 & 0.088 \\
        2.2 & $-0.30$ & 2.5 & 707.7 & 3.900 & 3.9240 & 0.097 \\
        2.3 & $-0.30$ & 1.7 & 658.8 & 3.912 & 3.9242 & 0.125 \\
        2.3 & $-0.30$ & 1.5 & 660.9 & 3.914 & 3.9243 & 0.137 \\
        \hline
    \end{tabular}
    \tablefoot{
    The models cover late main-sequence, turn-off, and early post-main-sequence phases.
    }
\end{table}

Considering the secondary component of the $\gamma$ Persei system, the best-fitting models cover a range of $\varepsilon\simeq 0.088 - 0.137$. 
However, the evolutionary phase cannot be uniquely determined by the $\varepsilon$ metric, as the best-fitting solutions cover late main-sequence, turn-off, and subgiant configurations (see Figure\,\ref{fig:best_fit_sec}). 
Table\,\ref{tab:secondary_best_models} collects the calculated stellar parameters for the four best-fitting models for this component.

\section{Discussion} \label{sec:discussion}

The discrepancy between the ages of the two stars suggests that the primary component is a rejuvenated merger product, which passed through the evolutionary phase commonly referred to as a blue straggler \citep[see, e.g., the recent review of][]{Schneider2025} and \cite{Schneideretal2025}. 
Blue stragglers are stars that appear younger than their neighboring stars in clusters \citep[e.g.,][]{Hellings1983,Podsiadlowskietal1992,Braunetal1995,Drayetal2007}.
Thus, their calculated age is an ``apparent age'' and originates from the replenishment of hydrogen in its core during the merger and their new location on the HRD.
\cite{Schneideretal2016} pointed out that the rejuvenation can be significant depending on the masses and the evolutionary stages of the merging components.
They showed that the rejuvenation can cover a range from at least a few per cent up to 80\% and more of the main-sequence lifetime of stars.
This scenario may apply to the primary component of the $\gamma$ Persei system.
While the primary component holds a rejuvenated age, the secondary component reflects the true evolutionary age of the system and acts as an anchor to determine the age of the system.

\vspace{1em}

Following this idea, based on the results of the stellar evolution models, we define the items listed below.

\begin{itemize}
    \item True age of the system ($T_{\rm true}$): inferred from the secondary component, yielding $\sim 750$--$900\,$Myr.
    \item Apparent age of the merger product ($t_{\rm app}$): inferred from the primary component, yielding $\sim 280$--$350\,$Myr.
\end{itemize}

We quantified the rejuvenation using an effective rejuvenation fraction of
\begin{equation}
R_{\rm eff} = 1 - \frac{t_{\rm app}}{T_{\rm true}},
\end{equation}
which directly measures the fractional reduction in apparent age relative to the true system age.
This quantity is an analog of the rejuvenation defined in terms of fractional main-sequence ages by \citet{Glebbeek2008} and \citet{Schneideretal2016}.

\subsection{Minimum rejuvenation requirements}

The minimum rejuvenation corresponds to the limiting case in which the merger occurred very recently, such that the present-day apparent age closely reflects the post-merger apparent age.
This represents a conservative lower limit as it neglects the short post-merger thermal-relaxation phase.
The latter can be done as the thermal relaxation is expected to occur on the Kelvin--Helmholtz timescale:
\begin{equation}
    t_{\rm KH} \simeq \frac{G M^2}{R L}.
\end{equation}
For the primary ($M \simeq 3.5\,M_\odot$, $R \simeq 22.7\,R_\odot$, $L \simeq 10^{2.45}\,L_\odot$),
this yields
\begin{equation}
    t_{\rm KH} \sim 5\times10^{4}\text{--}10^{5}\,\mathrm{yr},
\end{equation}
which is negligible compared to the inferred system age and the main-sequence evolutionary timescale.

The rejuvenation factor ($R$) can be constrained between the lower and upper bounds inferred from $T_{\rm true}$ and $t_{\rm app}$. These are listed as follows. 

\begin{itemize}
    \item Lower bound: $T_{\rm true} = 750\,\mathrm{Myr}$ and $t_{\rm app} = 350\,\mathrm{Myr}$,
    corresponding to the younger end of the calculated true age range and the older end of the inferred apparent age range,
    yielding
    \begin{equation}
        R_\mathrm{eff, low} = 1 - \frac{350}{750} \approx 0.53.
    \end{equation}

    \item Upper bound: $T_{\rm true} = 900\,\mathrm{Myr}$ and $t_{\rm app} = 280\,\mathrm{Myr}$,
    corresponding to the older end of the calculated true age range and the younger end of the inferred apparent age range,
    yielding
    \begin{equation}
        R_\mathrm{eff, up} = 1 - \frac{280}{900} \approx 0.69.
    \end{equation}
\end{itemize}

\noindent Thus, a rejuvenation of at least $50$--$70\%$ is required to explain the observed properties of the primary component with the system age inferred from the secondary component. 
The calculated range of the rejuvenation factor is consistent with the results of \citet{Schneideretal2016}, for example.

\subsection{Timing of the merger event}

Assuming a representative rejuvenation efficiency of $R = 0.8$, which corresponds to the approximate upper limit expected from main-sequence (or MS+MS) mergers \citep{Schneideretal2016}, we were able to constrain a window for the epoch of the merger ($T_{\rm merg}$). The apparent age of the rejuvenated star after the merger is reduced by a factor of $R_{\rm eff}$. 
Subsequently, after the thermal relaxation, the star evolves along the main sequence as a ``simple'', non-disturbed star.
The apparent, present-day age of the merged star is given by
\begin{equation}
    t_{\rm app} = T_{\rm true} - R_{\rm eff}\, T_{\rm merg},
\end{equation}
where $T_{\rm true}$ is the actual age of the system, $T_{\rm merg}$ is the time of the merger, and $t_{\rm app}$ is the apparent age calculated from observations.
Solving for $T_{\rm merg}$ yields

\begin{equation}
    T_{\rm merg} = \frac{T_{\rm true} - t_{\rm app}}{R_{\rm eff}}.
\end{equation}

Using the calculated ranges for the system age and apparent age of the primary component (see Table\,\ref{tab:mesa_results}), one can constrain a time window for the epoch of the merger as
\begin{equation}
    T_{\rm merg, min} = \frac{T_{\rm true, \ min} - t_{\rm app, \ max}}{R_{\rm eff}} \approx \frac{750 - 350}{0.8} \approx 500\,\mathrm{Myr},
\end{equation}
\begin{equation}
    T_{\rm merg, \ max} = \frac{T_{\rm true, \ max} - t_{\rm app, \ min}}{R_{\rm eff}} \approx \frac{900 - 280}{0.8} \approx 775\,\mathrm{Myr}.
\end{equation}
Thus, the merger must have occurred no later than $\sim 500$--$775\,\mathrm{Myr}$ ago.
Assuming that the system is $750-900$\,Myr old, the merger likely took place $\sim 150-200$\,Myr after formation, while the progenitor stars were still on the main sequence.

\subsection{Constraining the progenitor masses for the primary component}
Assuming that the present-day primary component was produced by a MS+MS merger, we can constrain the masses of the progenitor merging stars.
This can be obtained by using the current age of the system from the secondary component of the $\gamma$~Persei system and the estimated lifetime of the merging main-sequence stars ($t_\mathrm{MS}$). 
The latter can be approximated as
\begin{equation}
    t_{\rm MS} \sim \frac{M}{L},
\end{equation}
where $M$ is the stellar mass and $L$ the luminosity \citep[e.g.,][]{Kippenhahnbook2012}. 
Assuming the standard mass--luminosity relation on the main sequence:
\begin{equation}
    L \sim M^{3.5},
\end{equation}
the lifetime of the main sequence becomes
\begin{equation}
    t_{\rm MS} \sim 10^{10}\,{\rm yr} \left(\frac{M}{M_\odot}\right)^{-2.5}.
    \label{eq:ms_age}
\end{equation}

Assuming that the merger occurred recently, but sufficiently long ago for the remnant to have completed its thermal relaxation, the age of the system can be used to constrain the maximum masses of the progenitor stars. 
Given an estimated system age of $T_{\rm true} \simeq 750$--$900\,{\rm Myr}$, the corresponding main-sequence turn-off point mass ($M_\mathrm{TO}$) can be calculated from Equation\,\ref{eq:ms_age}, yielding
\begin{equation}
    M_{\rm TO} \simeq 2.3\text{--}2.5\,M_\odot.
\end{equation}
This provides an upper limit on the masses of the progenitor stars for the close binary.
Any star exceeding $M_{\rm TO}$ would have already left the main
sequence by the time of coalescence, preventing a MS+MS merger at the present epoch.

The present-day mass of the primary component is $M_1 = 3.5 \pm 0.3\,M_\odot$. 
Assuming a modest mass loss during the merger ($\eta \simeq 0.9$--$0.95$), the total mass of the progenitor binary is constrained by
\begin{equation}
\label{eq:prog}
    M_{1, a} + M_{1, b} = \frac{M_1}{\eta}
    \simeq 3.4\text{--}4.2\,M_\odot.
\end{equation}
As both progenitors must remain below the turn-off mass ($M_{\rm TO} \simeq 2.3$--$2.5\,M_\odot$), the two masses cannot vary independently. 

For the lowest allowed total mass ($M_{1,a}+M_{1,b}=3.4\,M_\odot$), the extreme case corresponds to a point that lies near the turn-off point, while the other one reaches the absolute minimum that fulfills Equation\,\ref{eq:prog} assuming that $M_\mathrm{1,b} \ge M_\mathrm{1,a}$:
\begin{equation}
\begin{split}
    M_{1,a}^\mathrm{lower} \simeq 0.9\text{--}1.1\,M_\odot,\\
    M_{1,b}^\mathrm{lower} \simeq 2.3\text{--}2.5\,M_\odot.
\end{split}
\end{equation}

Assuming the highest allowed total mass for the system ($M_{1,a}+M_{1,b}=4.2\,M_\odot$), we can constrain the progenitor masses with the turn-off point from its conservative upper limit as $M_{\rm TO} =2.5\,M_\odot$ constrained by $M_\mathrm{1,b} \ge M_\mathrm{1,a}$:
\begin{equation}
\begin{split}
    M_{1,a}^\mathrm{upper} \simeq 1.7\text{--}2.1\,M_\odot, \\
    M_{1,b}^\mathrm{upper} \simeq 2.1\text{--}2.5\,M_\odot.
\end{split}
\end{equation}

\begin{figure}
    \centering
    \includegraphics[width=\linewidth]{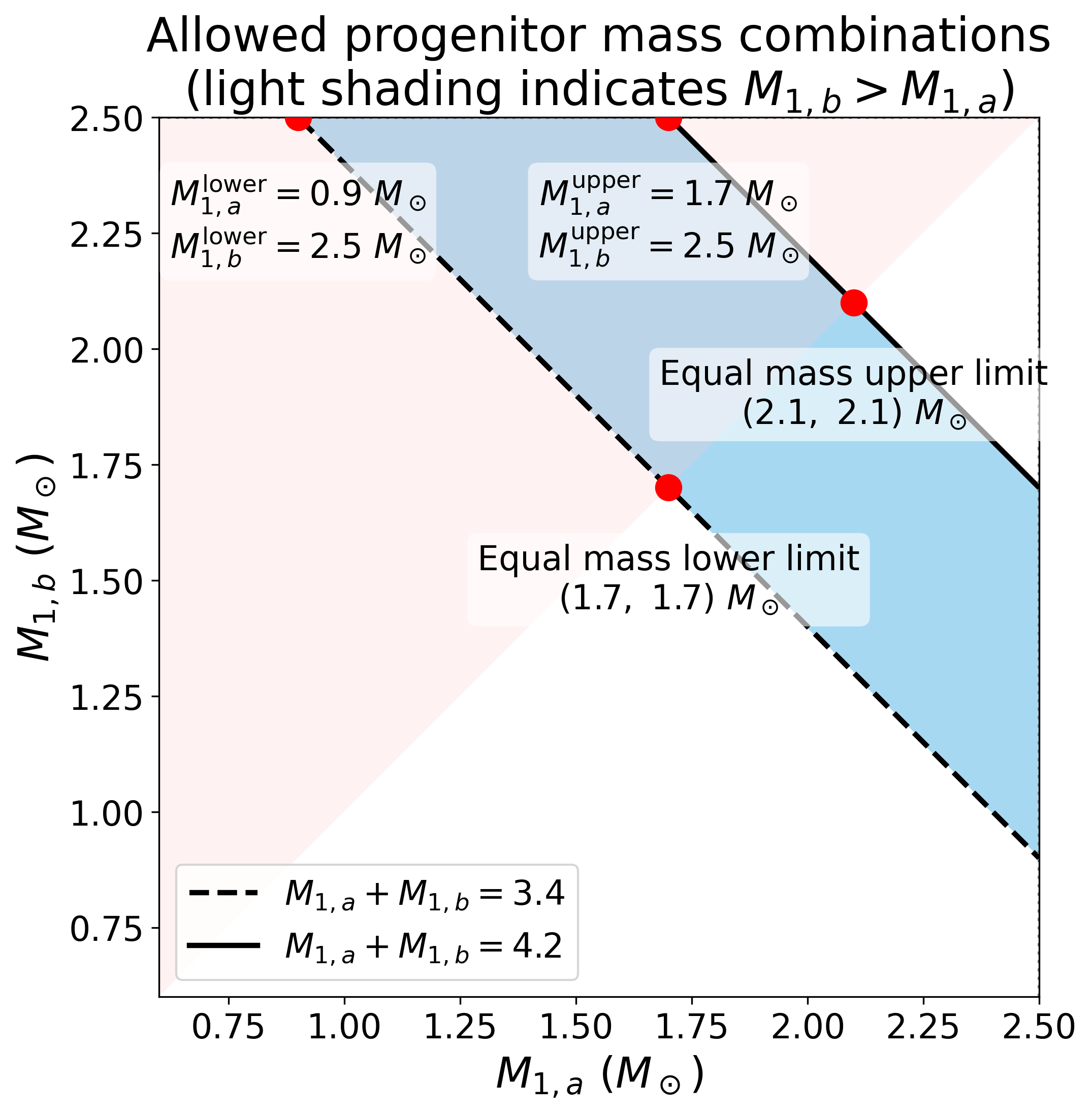}
    \caption{Allowed combinations of progenitor masses $(M_{1,a},M_{1,b})$. The blue region shows the parameter space permitted by the total-mass constraint ($3.4$--$4.2\,M_\odot$) and the requirement that both stars remain below the turn-off mass ($M_{\rm TO}\simeq2.3-2.5\,M_\odot$). The light red shaded region denotes the region where $M_{1,b}>M_{1,a}$, as assumed throughout the analysis. The red points indicate the extreme allowed configurations discussed in the text: the lower mass limit ($M_{1,a}^{\rm lower}=0.9\,M_\odot$, $M_{1,b}^{\rm lower}=2.5\,M_\odot$), the upper mass limit ($M_{1,a}^{\rm upper}=1.7\,M_\odot$, $M_{1,b}^{\rm upper}=2.5\,M_\odot$), and the equal-mass upper case ($M_{1,a}=M_{1,b}=2.1\,M_\odot$). Only mass pairs that lie within the diagonal band satisfy all physical constraints.}
    \label{fig:prog_band}
\end{figure}

The constraints in Equation~\ref{eq:prog}, together with the requirement that both progenitors remain below the turn-off mass, do not define two independent mass intervals for $(M_{1,a},M_{1,b})$. 
Instead, the allowed solutions occupy a narrow diagonal band in the $(M_{1,a},M_{1,b})$ plane, which is bounded by the total-mass limits ($3.4$--$4.2\,M_\odot$) and by the condition $M_{1,b}\ge M_{1,a}$. 
The resulting allowed parameter space of progenitor masses is shown in Fig.~\ref{fig:prog_band}. 
As illustrated in Fig.~\ref{fig:prog_band}, only mass pairs lying within this band are physically permitted.

We note that the merger efficiency is not fixed but depends on the evolutionary state, mass ratio, and internal structure of the progenitor stars. 
Hydrodynamic simulations show that MS+MS collisions typically eject only a few percent of the total mass, with higher mass loss occurring for more evolved or more equal-mass encounters \citep[see, e.g.,][]{Glebbeeketal2013}.

\subsection{Dynamical plausibility of a former triple configuration}

The proposed merger scenario suggests that the present-day $\gamma$~Persei system may have originated from an initially hierarchical triple system. 
Although a full dynamical reconstruction is beyond the scope of this paper, the basic timescales indicate that such a configuration can be estimated as follows.
Assuming that the primordial triple system had a sufficiently large mutual inclination, secular three-body evolution could have driven Lidov--Kozai oscillations, periodically increasing the eccentricity of the inner binary and thereby reducing its periastron distance. 
This process can lead to strong tidal dissipation, causing the merger of the inner, close binary stars. 
The characteristic Lidov--Kozai timescale is approximately

\begin{equation}
    t_{\rm Kozai} \sim \frac{P_{\rm out}^2}{P_{\rm in}} \frac{m_1+m_2+m_3}{m_3} (1-e_{\rm out}^2)^{3/2},
\end{equation}where $P_{\rm in}$ and $P_{\rm out}$ are the inner and outer orbital periods, and $m_1$, $m_2$, and $m_3$ are the component masses. Using the observed orbital period of the surviving binary, $P_{\rm out}=5329$\,d; its eccentricity, $e_{\rm out}=0.785$; and the present-day component masses within their observationally allowed ranges, we obtain Lidov--Kozai timescales on the order of \citep{Kozai1962,Lidov1962}

\begin{equation}
    t_{\rm Kozai} \sim 10^3\text{--}10^4\,{\rm yr}
,\end{equation}
assuming a representative compact primordial inner binary system with an orbital period of $P_{\rm in}\sim3$--$30$\,d.

We emphasize that the estimated Kozai–Lidov timescale is only an order‑of‑magnitude approximation to the relevant secular evolution, and the actual coalescence time of the inner binary need not be equal to $t_{\rm Kozai}$.
In practice, a stellar merger requires repeated eccentricity excitation over many cycles, but the characteristic secular timescale is still many orders of magnitude shorter than the inferred system age of $\sim750$--$900$\,Myr, showing that the proposed merger pathway is dynamically feasible. 
The currently large eccentricity of the surviving binary is also not inconsistent with this picture; on the contrary, an eccentric outer orbit may enhance secular eccentricity excitation in the primordial triple and thereby facilitate merger of the inner pair.

These considerations are consistent with previous studies of blue straggler formation in primordial triple systems. 
In particular, \citet{Perets2009} and \citet{Naots2014} showed that Kozai--Lidov–driven mergers of inner binaries naturally produce long-period, eccentric post-merger binaries, a characteristic outcome of triple-induced evolution. 
The present-day properties of $\gamma$~Persei therefore fit well within the scenario of triple-origin merger.

\section{Conclusions} \label{sec:conclusion}

In this work, we investigated whether the present-day configuration of the
$\gamma$~Persei system can be explained under the assumption of coeval formation and standard single-star evolution, using the observational constraints reported in the literature \citep[e.g.,][]{Griffinetal1994, Pourbaixetal2004, Diamantetal2023, AdammandMolnar2025, RosasPortilla2024} and in the companion paper presenting the asteroseismic analysis of the primary \citep{Adametal2026}.
Isochrone fitting based on the \texttt{MIST} database demonstrates that, although formal coeval solutions exist for certain metallicities, these solutions require stellar masses and chemical compositions that are inconsistent with the directly measured system parameters.
In particular, joint isochrone solutions systematically favor significantly lower masses and more metal-poor compositions than those that can be found in the literature.
We therefore conclude that \texttt{MIST}-based coeval solutions are astrophysically implausible for the $\gamma$~Persei system.

To further investigate the coevality of the system, we constructed a grid of stellar evolution models using \texttt{MESA}.
Across all the applied model parameters, we consistently found an evolutionary mismatch between the two components.
The primary component is found to be in a post-main-sequence evolutionary phase, i.e., on the red giant branch, or, most likely, in the red clump.
In contrast, the secondary component lies close to the turn-off point of the main sequence or at the very beginning of the subgiant branch. 

To overcome the age discrepancy, we conclude that the primary component of the system is a rejuvenated (former blue straggler) object formed through a merger of two main-sequence stars \citep[e.g.,][]{Glebbeek2008, Schneideretal2016, Schneider2025}.
In this scenario, i.e., in the present-day, the secondary component reflects the true age of the system, while the merger product appears significantly younger due to its placement on the HRD and the replenishment of hydrogen in its core. 

This suggests that the $\gamma$~Persei system formed as a triple-star system, where an inner, closer binary merged during the main-sequence phase, and the outer component evolved mainly undisturbed.
Using the evolutionary state of the non-merged star as an age anchor, we calculated the true age of the system to be $\sim 750$--$900$~Myr.
The merger is constrained to have occurred $\sim 500$--$775$~Myr after system formation, and the total mass of the progenitor binary is limited to $3.4$--$4.2\,M_\odot$. 

Furthermore, we conclude that the progenitor masses do not span independent intervals for $M_{1,a}$ and $M_{1,b}$, but must instead lie within the narrow diagonal band shown in Fig.~\ref{fig:prog_band}. 
Within this band, assuming that $M_{1,b} \ge M_{1,a}$, the primary component is constrained by the turn-off mass as an upper limit and $M_\mathrm{TO} / 2$ as a lower limit; $M_{1,b}$  falls within a mass range of $M_{1,b} \simeq 1.7$--$2.5\,M_\odot$. 
On the contrary, the lower mass companion lies within the $M_{1,a}\simeq0.9$--$2.1\,M_\odot$ range, depending on the total mass of the system.

At an age below 1~Gyr, the system is too young to infer an independent age constraint from the [Y/Mg] relative abundances, as this method is only accurate on longer timescales, following the chemical evolution of the Milky Way \citep{Berger-2022}. Further tests of the system would be achievable with more detailed asteroseismic observations of the primary to detect mixed modes that probe the core, and thus could differentiate between evolutionary stages, but those would require \textit{Kepler}-like coverage. Unfortunately, $\gamma$~Per does not fall into the proposed long observation fields of the PLATO mission that would provide such data \citep{Nascimbeni-2022}, but that mission could reveal further examples of post-merger systems.  

\begin{acknowledgements}
This research was supported by the `SeismoLab' KKP-137523 \'Elvonal grant of the Hungarian Research, Development and Innovation Office (NKFIH), and by the LP2025-14/2025 Lendület grant of the Hungarian Academy of Sciences. 

The simulations were performed on the high-performance computing machine at HUN-REN CSFK CSI acquired through the EC Horizon2020 project OPTICON (Grant Agreement No. 730890).

This research made use of NASA's Astrophysics Data System Bibliographic Services, as well as of the SIMBAD and VizieR databases operated at CDS, Strasbourg, France.

This paper was written with the assistance of a large language model (e.g., Gemini by Google, Microsoft Copilot, and ChatGPT by OpenAI) for linguistic  refinement. 
All content and any errors remain the sole responsibility of the first author.
\end{acknowledgements}


\bibliographystyle{aa} % style aa.bst
\bibliography{refs} 

@ARTICLE{AdammandMolnar2025,
       author = {{{\'A}d{\'a}m}, Roz{\'a}lia Z. and {Moln{\'a}r}, L{\'a}szl{\'o}},
        title = "{Filling the Gap: The Missing Eclipses of {\ensuremath{\gamma}} Persei from 2005 and from 2006}",
      journal = {\aj},
         year = 2025,
        month = apr,
       volume = {169},
       number = {4},
          eid = {196},
        pages = {196},
          doi = {10.3847/1538-3881/adb60d},
archivePrefix = {arXiv},
       eprint = {2503.00573},
 primaryClass = {astro-ph.SR},
       adsurl = {https://ui.adsabs.harvard.edu/abs/2025AJ....169..196A}
}

@ARTICLE{Adametal2026,
       author = {
           {{\'A}d{\'a}m}, Roz{\'a}lia Z. and
           {Moln{\'a}r}, L{\'a}szl{\'o} and
           {Szab{\'o}}, R{\'o}bert and
           {Kalup}, Csilla and
           {Grundahl}, Frank and
           {Huber}, Daniel and
           {Fredslund}, Mads Skakke and
           {Pall{\'e}}, Pere L. and
           {Tarczay-Neh{\'e}z}, D{\'o}ra
       },
        title = "{Asteroseismic Analysis of the Merger Product Red Giant in the γ Persei System}",
      journal = {\aap},
         year = 2026,
         note = {submitted}
}

@PROCEEDINGS{AkeandGriffin2015,
        title = "{Giants of Eclipse: The {\ensuremath{\zeta}} Aurigae Stars and Other Binary Systems}",
    booktitle = {Giants of Eclipse: The {\ensuremath{\zeta}} Aurigae Stars and Other Binary Systems},
         year = 2015,
       editor = {{Ake}, Thomas B. and {Griffin}, Elizabeth},
       series = {Astrophysics and Space Science Library},
       volume = {408},
        month = jan,
          doi = {10.1007/978-3-319-09198-3},
       adsurl = {https://ui.adsabs.harvard.edu/abs/2015ASSL..408.....A}
}

@article{Angulo1999,
        title = {A compilation of charged-particle induced thermonuclear reaction rates},
        journal = {Nuclear Physics A},
        volume = {656},
        number = {1},
        pages = {3-183},
        year = {1999},
        issn = {0375-9474},
        doi = {https://doi.org/10.1016/S0375-9474(99)00030-5},
        url = {https://www.sciencedirect.com/science/article/pii/S0375947499000305},
        author = {C. Angulo and M. Arnould and M. Rayet and P. Descouvemont and D. Baye and C. Leclercq-Willain and A. Coc and S. Barhoumi and P. Aguer and C. Rolfs and R. Kunz and J.W. Hammer and A. Mayer and T. Paradellis and S. Kossionides and C. Chronidou and K. Spyrou and S. Degl'Innocenti and G. Fiorentini and B. Ricci and S. Zavatarelli and C. Providencia and H. Wolters and J. Soares and C. Grama and J. Rahighi and A. Shotter and M. {Lamehi Rachti}}
}

@ARTICLE{Asplundetal2009,
       author = {{Asplund}, Martin and {Grevesse}, Nicolas and {Sauval}, A. Jacques and {Scott}, Pat},
        title = "{The Chemical Composition of the Sun}",
      journal = {\araa},
         year = 2009,
        month = sep,
       volume = {47},
       number = {1},
        pages = {481-522},
          doi = {10.1146/annurev.astro.46.060407.145222},
archivePrefix = {arXiv},
       eprint = {0909.0948},
 primaryClass = {astro-ph.SR},
       adsurl = {https://ui.adsabs.harvard.edu/abs/2009ARA&A..47..481A}
}

@ARTICLE{Braunetal1995,
       author = {{Braun}, H. and {Langer}, N.},
        title = "{Effects of accretion onto massive main sequence stars.}",
      journal = {\aap},
         year = 1995,
        month = may,
       volume = {297},
        pages = {483},
       adsurl = {https://ui.adsabs.harvard.edu/abs/1995A&A...297..483B}
}

@ARTICLE{Campbell1908,
       author = {{Campbell}, W.~W.},
        title = "{Eighteen Stars Whose Radial Velocities Vary}",
      journal = {\pasp},
         year = 1908,
        month = dec,
       volume = {20},
       number = {123},
        pages = {293},
          doi = {10.1086/121853},
       adsurl = {https://ui.adsabs.harvard.edu/abs/1908PASP...20..293C}
}

@ARTICLE{Campbelletal1909,
       author = {{Campbell}, W.~W.},
        title = "{Eleven stars having variable radial velocities.}",
      journal = {\apj},
         year = 1909,
        month = apr,
       volume = {29},
        pages = {224-228},
          doi = {10.1086/141644},
       adsurl = {https://ui.adsabs.harvard.edu/abs/1909ApJ....29..224C}
}

@ARTICLE{Cyburt2010,
       author = {{Cyburt}, Richard H. and {Amthor}, A. Matthew and {Ferguson}, Ryan and {Meisel}, Zach and {Smith}, Karl and {Warren}, Scott and {Heger}, Alexander and {Hoffman}, R.~D. and {Rauscher}, Thomas and {Sakharuk}, Alexander and {Schatz}, Hendrik and {Thielemann}, F.~K. and {Wiescher}, Michael},
        title = "{The JINA REACLIB Database: Its Recent Updates and Impact on Type-I X-ray Bursts}",
      journal = {\apjs},
         year = 2010,
        month = jul,
       volume = {189},
       number = {1},
        pages = {240-252},
          doi = {10.1088/0067-0049/189/1/240},
       adsurl = {https://ui.adsabs.harvard.edu/abs/2010ApJS..189..240C}
}

@ARTICLE{Diamantetal2023,
       author = {{Diamant}, S.~J.~M. and {Schr{\"o}der}, K. -P. and {Jack}, D. and {Rosas-Portilla}, F. and {Fridlund}, M. and {Schmitt}, J.~H.~M.~M.},
        title = "{Discovery of an extended G giant chromosphere in the 2019 eclipse of {\ensuremath{\gamma}} Per}",
      journal = {\aap},
         year = 2023,
        month = jun,
       volume = {674},
          eid = {A162},
        pages = {A162},
          doi = {10.1051/0004-6361/202245241},
       adsurl = {https://ui.adsabs.harvard.edu/abs/2023A&A...674A.162D}
}

@ARTICLE{Drayetal2007,
       author = {{Dray}, Lynnette M. and {Tout}, Christopher A.},
        title = "{On rejuvenation in massive binary systems}",
      journal = {\mnras},
         year = 2007,
        month = mar,
       volume = {376},
       number = {1},
        pages = {61-70},
          doi = {10.1111/j.1365-2966.2007.11431.x},
archivePrefix = {arXiv},
       eprint = {astro-ph/0612539},
 primaryClass = {astro-ph},
       adsurl = {https://ui.adsabs.harvard.edu/abs/2007MNRAS.376...61D}
}

@ARTICLE{Fergusonetal2005,
       author = {{Ferguson}, Jason W. and {Alexander}, David R. and {Allard}, France and {Barman}, Travis and {Bodnarik}, Julia G. and {Hauschildt}, Peter H. and {Heffner-Wong}, Amanda and {Tamanai}, Akemi},
        title = "{Low-Temperature Opacities}",
      journal = {\apj},
         year = 2005,
        month = apr,
       volume = {623},
       number = {1},
        pages = {585-596},
          doi = {10.1086/428642},
archivePrefix = {arXiv},
       eprint = {astro-ph/0502045},
 primaryClass = {astro-ph},
       adsurl = {https://ui.adsabs.harvard.edu/abs/2005ApJ...623..585F}
}

@dataset{GaiaEDR3,
       author = {{Gaia Collaboration}},
        title = "{VizieR Online Data Catalog: Gaia EDR3 (Gaia Collaboration, 2020)}",
 howpublished = {VizieR On-line Data Catalog: I/350.  Originally published in: 2021A\&A...649A...1G},
         year = 2020,
        month = nov,
          eid = {I/350},
          doi = {10.26093/cds/vizier.1350},
       adsurl = {https://ui.adsabs.harvard.edu/abs/2020yCat.1350....0G}
}

@ARTICLE{Ginestet2002,
       author = {{Ginestet}, N. and {Carquillat}, J.~M.},
        title = "{Spectral Classification of the Hot Components of a Large Sample of Stars with Composite Spectra, and Implication for the Absolute Magnitudes of the Cool Supergiant Components.}",
      journal = {\apjs},
         year = 2002,
        month = dec,
       volume = {143},
       number = {2},
        pages = {513-537},
          doi = {10.1086/342942},
       adsurl = {https://ui.adsabs.harvard.edu/abs/2002ApJS..143..513G}
}

@ARTICLE{Glebbeek2008,
       author = {{Glebbeek}, E. and {Pols}, O.~R.},
        title = "{Evolution of stellar collision products in open clusters. II. A grid of low-mass collisions}",
      journal = {\aap},
         year = 2008,
        month = sep,
       volume = {488},
       number = {3},
        pages = {1017-1025},
          doi = {10.1051/0004-6361:200809931},
archivePrefix = {arXiv},
       eprint = {0806.0865},
 primaryClass = {astro-ph},
       adsurl = {https://ui.adsabs.harvard.edu/abs/2008A&A...488.1017G}
}

@article{Glebbeeketal2013,
    author = {Glebbeek, Evert and Gaburov, Evghenii and Portegies Zwart, Simon and Pols, Onno R.},
    title = {Structure and evolution of high-mass stellar mergers},
    journal = {Monthly Notices of the Royal Astronomical Society},
    volume = {434},
    number = {4},
    pages = {3497-3510},
    year = {2013},
    month = {08},
    issn = {0035-8711},
    doi = {10.1093/mnras/stt1268},
    url = {https://doi.org/10.1093/mnras/stt1268},
    eprint = {https://academic.oup.com/mnras/article-pdf/434/4/3497/18501344/stt1268.pdf},
}

@ARTICLE{Griffinetal1994,
       author = {{Griffin}, R.~F. and {Griffin}, R.~E.~M. and {Snyder}, L.~F. and {Schroder}, K. -P. and {Pray}, D. and {Ohshima}, O. and {Tokoro}, T. and {Clark}, W.~E. and {Williams}, D.~B. and {Houchen}, M.~B. and {Arai}, K. and {Krisciunas}, K. and {Watson}, J.},
        title = "{The Eclipse of Gamma Persei}",
      journal = {International Amateur-Professional Photoelectric Photometry Communications},
         year = 1994,
        month = sep,
       volume = {57},
        pages = {31},
       adsurl = {https://ui.adsabs.harvard.edu/abs/1994IAPPP..57...31G}
}

@INPROCEEDINGS{Griffin2007,
       author = {{Griffin}, R. Elizabeth},
        title = "{{\ensuremath{\gamma}} Per: Bright, but Ill-Understood}",
    booktitle = {Binary Stars as Critical Tools \& Tests in Contemporary Astrophysics},
         year = 2007,
       editor = {{Hartkopf}, William I. and {Harmanec}, Petr and {Guinan}, Edward F.},
       series = {IAU Symposium},
       volume = {240},
        month = aug,
        pages = {645-649},
          doi = {10.1017/S1743921307006151},
       adsurl = {https://ui.adsabs.harvard.edu/abs/2007IAUS..240..645G}
}

@ARTICLE{Hartkopf2001,
       author = {{Hartkopf}, William I. and {Mason}, Brian D. and {Worley}, Charles E.},
        title = "{The 2001 US Naval Observatory Double Star CD-ROM. II. The Fifth Catalog of Orbits of Visual Binary Stars}",
      journal = {\aj},
         year = 2001,
        month = dec,
       volume = {122},
       number = {6},
        pages = {3472-3479},
          doi = {10.1086/323921},
       adsurl = {https://ui.adsabs.harvard.edu/abs/2001AJ....122.3472H}
}

@ARTICLE{Hellings1983,
       author = {{Hellings}, P.},
        title = "{Phenomenological Study of Massive Accretion Stars}",
      journal = {\apss},
         year = 1983,
        month = oct,
       volume = {96},
       number = {1},
        pages = {37-54},
          doi = {10.1007/BF00661941},
       adsurl = {https://ui.adsabs.harvard.edu/abs/1983Ap&SS..96...37H}
}

@ARTICLE{IglesiasandRogers1993,
       author = {{Iglesias}, Carlos A. and {Rogers}, Forrest J.},
        title = "{Radiative Opacities for Carbon- and Oxygen-rich Mixtures}",
      journal = {\apj},
         year = 1993,
        month = aug,
       volume = {412},
        pages = {752},
          doi = {10.1086/172958},
       adsurl = {https://ui.adsabs.harvard.edu/abs/1993ApJ...412..752I}
}

@ARTICLE{Iglesiasetal1996,
       author = {{Iglesias}, Carlos A. and {Rogers}, Forrest J.},
        title = "{Updated Opal Opacities}",
      journal = {\apj},
         year = 1996,
        month = jun,
       volume = {464},
        pages = {943},
          doi = {10.1086/177381},
       adsurl = {https://ui.adsabs.harvard.edu/abs/1996ApJ...464..943I}
}

@ARTICLE{Jermyn2023,
       author = {{Jermyn}, Adam S. and {Bauer}, Evan B. and {Schwab}, Josiah and {Farmer}, R. and {Ball}, Warrick H. and {Bellinger}, Earl P. and {Dotter}, Aaron and {Joyce}, Meridith and {Marchant}, Pablo and {Mombarg}, Joey S.~G. and {Wolf}, William M. and {Sunny Wong}, Tin Long and {Cinquegrana}, Giulia C. and {Farrell}, Eoin and {Smolec}, R. and {Thoul}, Anne and {Cantiello}, Matteo and {Herwig}, Falk and {Toloza}, Odette and {Bildsten}, Lars and {Townsend}, Richard H.~D. and {Timmes}, F.~X.},
        title = "{Modules for Experiments in Stellar Astrophysics (MESA): Time-dependent Convection, Energy Conservation, Automatic Differentiation, and Infrastructure}",
      journal = {\apjs},
         year = 2023,
        month = mar,
       volume = {265},
       number = {1},
          eid = {15},
        pages = {15},
          doi = {10.3847/1538-4365/acae8d},
archivePrefix = {arXiv},
       eprint = {2208.03651},
 primaryClass = {astro-ph.SR},
       adsurl = {https://ui.adsabs.harvard.edu/abs/2023ApJS..265...15J}
}

@ARTICLE{JoyceandTayar2023,
       author = {{Joyce}, Meridith and {Tayar}, Jamie},
        title = "{A Review of the Mixing Length Theory of Convection in 1D Stellar Modeling}",
      journal = {Galaxies},
         year = 2023,
        month = jun,
       volume = {11},
       number = {3},
          eid = {75},
        pages = {75},
          doi = {10.3390/galaxies11030075},
archivePrefix = {arXiv},
       eprint = {2303.09596},
 primaryClass = {astro-ph.SR},
       adsurl = {https://ui.adsabs.harvard.edu/abs/2023Galax..11...75J}
}

@ARTICLE{Kozai1962,
       author = {{Kozai}, Yoshihide},
        title = "{Secular perturbations of asteroids with high inclination and eccentricity}",
      journal = {\aj},
         year = 1962,
        month = nov,
       volume = {67},
        pages = {591-598},
          doi = {10.1086/108790},
       adsurl = {https://ui.adsabs.harvard.edu/abs/1962AJ.....67..591K}
}

@ARTICLE{LiandJoyce2025,
       author = {{Li}, Yaguang and {Joyce}, Meridith},
        title = "{Beyond MESA Defaults: The Impact of Structural Resolution Uncertainty in p-mode Asteroseismology}",
      journal = {\apj},
         year = 2025,
        month = nov,
       volume = {994},
       number = {1},
          eid = {127},
        pages = {127},
          doi = {10.3847/1538-4357/ae0c1a},
archivePrefix = {arXiv},
       eprint = {2501.13207},
 primaryClass = {astro-ph.SR},
       adsurl = {https://ui.adsabs.harvard.edu/abs/2025ApJ...994..127L}
}

@ARTICLE{Lidov1962,
       author = {{Lidov}, M.~L.},
        title = "{The evolution of orbits of artificial satellites of planets under the action of gravitational perturbations of external bodies}",
      journal = {\planss},
         year = 1962,
        month = oct,
       volume = {9},
       number = {10},
        pages = {719-759},
          doi = {10.1016/0032-0633(62)90129-0},
       adsurl = {https://ui.adsabs.harvard.edu/abs/1962P&SS....9..719L}
}

@ARTICLE{Mauryetal1897,
       author = {{Maury}, Antonia C. and {Pickering}, Edward C.},
        title = "{Spectra of bright stars photographed with the 11-inch Draper Telescope as part of the Henry Draper Memorial.}",
      journal = {Annals of Harvard College Observatory},
         year = 1897,
        month = jan,
       volume = {28},
        pages = {1-128},
       adsurl = {https://ui.adsabs.harvard.edu/abs/1897AnHar..28....1M}
}

@ARTICLE{McLaughlin1948,
       author = {{McLaughlin}, Dean B.},
        title = "{The orbit of gamma Persei.}",
      journal = {\aj},
         year = 1948,
        month = jan,
       volume = {53},
        pages = {200},
          doi = {10.1086/106121},
       adsurl = {https://ui.adsabs.harvard.edu/abs/1948AJ.....53Q.200M}
}

@ARTICLE{McWilliam1990,
       author = {{McWilliam}, Andrew},
        title = "{High-Resolution Spectroscopic Survey of 671 GK Giants. I. Stellar Atmosphere Parameters and Abundances}",
      journal = {\apjs},
         year = 1990,
        month = dec,
       volume = {74},
        pages = {1075},
          doi = {10.1086/191527},
       adsurl = {https://ui.adsabs.harvard.edu/abs/1990ApJS...74.1075M}
}

@ARTICLE{MIST0,
       author = {{Dotter}, Aaron},
        title = "{MESA Isochrones and Stellar Tracks (MIST) 0: Methods for the Construction of Stellar Isochrones}",
      journal = {\apjs},
         year = 2016,
        month = jan,
       volume = {222},
       number = {1},
          eid = {8},
        pages = {8},
          doi = {10.3847/0067-0049/222/1/8},
archivePrefix = {arXiv},
       eprint = {1601.05144},
 primaryClass = {astro-ph.SR},
       adsurl = {https://ui.adsabs.harvard.edu/abs/2016ApJS..222....8D}
}

@ARTICLE{MIST1,
       author = {{Choi}, Jieun and {Dotter}, Aaron and {Conroy}, Charlie and {Cantiello}, Matteo and {Paxton}, Bill and {Johnson}, Benjamin D.},
        title = "{Mesa Isochrones and Stellar Tracks (MIST). I. Solar-scaled Models}",
      journal = {\apj},
         year = 2016,
        month = jun,
       volume = {823},
       number = {2},
          eid = {102},
        pages = {102},
          doi = {10.3847/0004-637X/823/2/102},
archivePrefix = {arXiv},
       eprint = {1604.08592},
 primaryClass = {astro-ph.SR},
       adsurl = {https://ui.adsabs.harvard.edu/abs/2016ApJ...823..102C}
}

@ARTICLE{Naots2014,
       author = {{Naoz}, Smadar and {Fabrycky}, Daniel C.},
        title = "{Mergers and Obliquities in Stellar Triples}",
      journal = {\apj},
         year = 2014,
        month = oct,
       volume = {793},
       number = {2},
          eid = {137},
        pages = {137},
          doi = {10.1088/0004-637X/793/2/137},
archivePrefix = {arXiv},
       eprint = {1405.5223},
 primaryClass = {astro-ph.SR},
       adsurl = {https://ui.adsabs.harvard.edu/abs/2014ApJ...793..137N}
}

@ARTICLE{Paxton2011,
       author = {{Paxton}, Bill and {Bildsten}, Lars and {Dotter}, Aaron and {Herwig}, Falk and {Lesaffre}, Pierre and {Timmes}, Frank},
        title = "{Modules for Experiments in Stellar Astrophysics (MESA)}",
      journal = {\apjs},
         year = 2011,
        month = jan,
       volume = {192},
       number = {1},
          eid = {3},
        pages = {3},
          doi = {10.1088/0067-0049/192/1/3},
archivePrefix = {arXiv},
       eprint = {1009.1622},
 primaryClass = {astro-ph.SR},
       adsurl = {https://ui.adsabs.harvard.edu/abs/2011ApJS..192....3P}
}

@ARTICLE{Paxton2013,
       author = {{Paxton}, Bill and {Cantiello}, Matteo and {Arras}, Phil and {Bildsten}, Lars and {Brown}, Edward F. and {Dotter}, Aaron and {Mankovich}, Christopher and {Montgomery}, M.~H. and {Stello}, Dennis and {Timmes}, F.~X. and {Townsend}, Richard},
        title = "{Modules for Experiments in Stellar Astrophysics (MESA): Planets, Oscillations, Rotation, and Massive Stars}",
      journal = {\apjs},
         year = 2013,
        month = sep,
       volume = {208},
       number = {1},
          eid = {4},
        pages = {4},
          doi = {10.1088/0067-0049/208/1/4},
archivePrefix = {arXiv},
       eprint = {1301.0319},
 primaryClass = {astro-ph.SR},
       adsurl = {https://ui.adsabs.harvard.edu/abs/2013ApJS..208....4P}
}

@ARTICLE{Paxton2015,
       author = {{Paxton}, Bill and {Marchant}, Pablo and {Schwab}, Josiah and {Bauer}, Evan B. and {Bildsten}, Lars and {Cantiello}, Matteo and {Dessart}, Luc and {Farmer}, R. and {Hu}, H. and {Langer}, N. and {Townsend}, R.~H.~D. and {Townsley}, Dean M. and {Timmes}, F.~X.},
        title = "{Modules for Experiments in Stellar Astrophysics (MESA): Binaries, Pulsations, and Explosions}",
      journal = {\apjs},
         year = 2015,
        month = sep,
       volume = {220},
       number = {1},
          eid = {15},
        pages = {15},
          doi = {10.1088/0067-0049/220/1/15},
archivePrefix = {arXiv},
       eprint = {1506.03146},
 primaryClass = {astro-ph.SR},
       adsurl = {https://ui.adsabs.harvard.edu/abs/2015ApJS..220...15P}
}

@ARTICLE{Paxton2018,
       author = {{Paxton}, Bill and {Schwab}, Josiah and {Bauer}, Evan B. and {Bildsten}, Lars and {Blinnikov}, Sergei and {Duffell}, Paul and {Farmer}, R. and {Goldberg}, Jared A. and {Marchant}, Pablo and {Sorokina}, Elena and {Thoul}, Anne and {Townsend}, Richard H.~D. and {Timmes}, F.~X.},
        title = "{Modules for Experiments in Stellar Astrophysics (MESA): Convective Boundaries, Element Diffusion, and Massive Star Explosions}",
      journal = {\apjs},
         year = 2018,
        month = feb,
       volume = {234},
       number = {2},
          eid = {34},
        pages = {34},
          doi = {10.3847/1538-4365/aaa5a8},
archivePrefix = {arXiv},
       eprint = {1710.08424},
 primaryClass = {astro-ph.SR},
       adsurl = {https://ui.adsabs.harvard.edu/abs/2018ApJS..234...34P}
}

@ARTICLE{Paxton2019,
       author = {{Paxton}, Bill and {Smolec}, R. and {Schwab}, Josiah and {Gautschy}, A. and {Bildsten}, Lars and {Cantiello}, Matteo and {Dotter}, Aaron and {Farmer}, R. and {Goldberg}, Jared A. and {Jermyn}, Adam S. and {Kanbur}, S.~M. and {Marchant}, Pablo and {Thoul}, Anne and {Townsend}, Richard H.~D. and {Wolf}, William M. and {Zhang}, Michael and {Timmes}, F.~X.},
        title = "{Modules for Experiments in Stellar Astrophysics (MESA): Pulsating Variable Stars, Rotation, Convective Boundaries, and Energy Conservation}",
      journal = {\apjs},
         year = 2019,
        month = jul,
       volume = {243},
       number = {1},
          eid = {10},
        pages = {10},
          doi = {10.3847/1538-4365/ab2241},
archivePrefix = {arXiv},
       eprint = {1903.01426},
 primaryClass = {astro-ph.SR},
       adsurl = {https://ui.adsabs.harvard.edu/abs/2019ApJS..243...10P}
}

@ARTICLE{Perets2009,
       author = {{Perets}, Hagai B. and {Fabrycky}, Daniel C.},
        title = "{On the Triple Origin of Blue Stragglers}",
      journal = {\apj},
         year = 2009,
        month = jun,
       volume = {697},
       number = {2},
        pages = {1048-1056},
          doi = {10.1088/0004-637X/697/2/1048},
archivePrefix = {arXiv},
       eprint = {0901.4328},
 primaryClass = {astro-ph.SR},
       adsurl = {https://ui.adsabs.harvard.edu/abs/2009ApJ...697.1048P}
}

@article{Piccottietal2020,
    author = {Piccotti, Luca and Docobo, Jose Angel and Carini, Roberta and Tamazian, Vakhtang S and Brocato, Enzo and Andrade, Manuel and Campo, Pedro P},
    title = {A study of the physical properties of SB2s with both the visual and spectroscopic orbits},
    journal = {\mnras},
    volume = {492},
    number = {2},
    pages = {2709-2721},
    year = {2020},
    month = {01},
    issn = {0035-8711},
    doi = {10.1093/mnras/stz3616},
    url = {https://doi.org/10.1093/mnras/stz3616},
    eprint = {https://academic.oup.com/mnras/article-pdf/492/2/2709/31876657/stz3616.pdf},
}

@ARTICLE{Pourbaixetal1999,
       author = {{Pourbaix}, D.},
        title = "{Gamma Persei: a challenge for stellar evolution models}",
      journal = {\aap},
         year = 1999,
        month = aug,
       volume = {348},
        pages = {127-132},
       adsurl = {https://ui.adsabs.harvard.edu/abs/1999A&A...348..127P}
}

@ARTICLE{Pourbaixetal2000,
       author = {{Pourbaix}, D.},
        title = "{Resolved double-lined spectroscopic binaries: A neglected source of hypothesis-free parallaxes and stellar masses}",
      journal = {\aaps},
         year = 2000,
        month = aug,
       volume = {145},
        pages = {215-222},
          doi = {10.1051/aas:2000237},
       adsurl = {https://ui.adsabs.harvard.edu/abs/2000A&AS..145..215P}
}

@ARTICLE{Pourbaixetal2004,
       author = {{Pourbaix}, D. and {Tokovinin}, A.~A. and {Batten}, A.~H. and {Fekel}, F.~C. and {Hartkopf}, W.~I. and {Levato}, H. and {Morrell}, N.~I. and {Torres}, G. and {Udry}, S.},
        title = "{S$_{B$^{9}$}$: The ninth catalogue of spectroscopic binary orbits}",
      journal = {\aap},
         year = 2004,
        month = sep,
       volume = {424},
        pages = {727-732},
          doi = {10.1051/0004-6361:20041213},
archivePrefix = {arXiv},
       eprint = {astro-ph/0406573},
 primaryClass = {astro-ph},
       adsurl = {https://ui.adsabs.harvard.edu/abs/2004A&A...424..727P}
}

@ARTICLE{Podsiadlowskietal1992,
       author = {{Podsiadlowski}, Ph. and {Joss}, P.~C. and {Hsu}, J.~J.~L.},
        title = "{Presupernova Evolution in Massive Interacting Binaries}",
      journal = {\apj},
         year = 1992,
        month = may,
       volume = {391},
        pages = {246},
          doi = {10.1086/171341},
       adsurl = {https://ui.adsabs.harvard.edu/abs/1992ApJ...391..246P}
}

@ARTICLE{RosasPortilla2024,
       author = {{Rosas-Portilla}, F. and {Jack}, D. and {Schr{\"o}der}, K. -P.},
        title = "{In pursuit of precise Ca II H\&K chromospheric surface fluxes: A gravity and temperature dependence}",
      journal = {\aap},
         year = 2024,
        month = dec,
       volume = {692},
          eid = {A189},
        pages = {A189},
          doi = {10.1051/0004-6361/202450691},
       adsurl = {https://ui.adsabs.harvard.edu/abs/2024A&A...692A.189R}
}

@ARTICLE{Schneideretal2016,
       author = {{Schneider}, F.~R.~N. and {Podsiadlowski}, Ph. and {Langer}, N. and {Castro}, N. and {Fossati}, L.},
        title = "{Rejuvenation of stellar mergers and the origin of magnetic fields in massive stars}",
      journal = {\mnras},
         year = 2016,
        month = apr,
       volume = {457},
       number = {3},
        pages = {2355-2365},
          doi = {10.1093/mnras/stw148},
archivePrefix = {arXiv},
       eprint = {1601.05084},
 primaryClass = {astro-ph.SR},
       adsurl = {https://ui.adsabs.harvard.edu/abs/2016MNRAS.457.2355S}
}

@ARTICLE{Schneider2025,
       author = {{Schneider}, Fabian R.~N.},
        title = "{Theory, Simulations and Observations of Stellar Mergers}",
      journal = {arXiv e-prints},
         year = 2025,
        month = sep,
          eid = {arXiv:2509.18421},
        pages = {arXiv:2509.18421},
          doi = {10.48550/arXiv.2509.18421},
archivePrefix = {arXiv},
       eprint = {2509.18421},
 primaryClass = {astro-ph.SR},
       adsurl = {https://ui.adsabs.harvard.edu/abs/2025arXiv250918421S}
}

@ARTICLE{Schneideretal2025,
       author = {{Schneider}, Fabian R.~N. and {Lau}, Mike Y.~M. and {Roepke}, Friedrich K.},
        title = "{Stellar mergers and common-envelope evolution}",
      journal = {arXiv e-prints},
         year = 2025,
        month = jan,
          eid = {arXiv:2502.00111},
        pages = {arXiv:2502.00111},
          doi = {10.48550/arXiv.2502.00111},
archivePrefix = {arXiv},
       eprint = {2502.00111},
 primaryClass = {astro-ph.SR},
       adsurl = {https://ui.adsabs.harvard.edu/abs/2025arXiv250200111S}
}

@article{TND2026,
doi = {10.1088/1538-3873/ae50b9},
url = {https://doi.org/10.1088/1538-3873/ae50b9},
year = {2026},
month = {apr},
publisher = {The Astronomical Society of the Pacific},
volume = {138},
number = {4},
pages = {044203},
author = {Tarczay-Nehéz, Dóra and Molnár, László and Joyce, Meridith},
title = {Stranger Things: A Grid-based Survey of Strange Modes in Post-Main Sequence Models},
journal = {Publications of the Astronomical Society of the Pacific},

}

@ARTICLE{XuandLi2004a,
       author = {{Xu}, H.~Y. and {Li}, Y.},
        title = "{Blue loops of intermediate mass stars . II. Metallicity and blue loops}",
      journal = {\aap},
         year = 2004,
        month = apr,
       volume = {418},
        pages = {225-233},
          doi = {10.1051/0004-6361:20040023},
       adsurl = {https://ui.adsabs.harvard.edu/abs/2004A&A...418..225X}
}

@ARTICLE{XuandLi2004b,
       author = {{Xu}, H.~Y. and {Li}, Y.},
        title = "{Blue loops of intermediate mass stars . I. CNO cycles and blue loops}",
      journal = {\aap},
         year = 2004,
        month = apr,
       volume = {418},
        pages = {213-224},
          doi = {10.1051/0004-6361:20040024},
       adsurl = {https://ui.adsabs.harvard.edu/abs/2004A&A...418..213X}
}

@ARTICLE{Ziolkowska2024,
       author = {{Zi{\'o}{\l}kowska}, O. and {Smolec}, R. and {Thoul}, A. and {Farrell}, E. and {Rathour}, R. Singh and {Hocd{\'e}}, V.},
        title = "{Toward a Comprehensive Grid of Cepheid Models with MESA. I. Uncertainties of the Evolutionary Tracks of Intermediate-mass Stars}",
      journal = {\apjs},
         year = 2024,
        month = oct,
       volume = {274},
       number = {2},
          eid = {30},
        pages = {30},
          doi = {10.3847/1538-4365/ad614d},
archivePrefix = {arXiv},
       eprint = {2408.07136},
 primaryClass = {astro-ph.SR},
       adsurl = {https://ui.adsabs.harvard.edu/abs/2024ApJS..274...30Z}
}

@Article{matplotlib,
  Author    = {Hunter, J. D.},
  Title     = {Matplotlib: A 2D graphics environment},
  Journal   = {Computing in Science \& Engineering},
  Volume    = {9},
  Number    = {3},
  Pages     = {90--95},
  publisher = {IEEE COMPUTER SOC},
  doi       = {10.1109/MCSE.2007.55},
  year      = 2007
}

@ARTICLE{numpy,
       author = {{van der Walt}, St{\'e}fan and {Colbert}, S. Chris and
         {Varoquaux}, Ga{\"e}l},
        title = "{The NumPy Array: A Structure for Efficient Numerical Computation}",
      journal = {Computing in Science and Engineering},
         year = "2011",
        month = "Mar",
       volume = {13},
       number = {2},
        pages = {22-30},
          doi = {10.1109/MCSE.2011.37},
archivePrefix = {arXiv},
       eprint = {1102.1523},
 primaryClass = {cs.MS},
       adsurl = {https://ui.adsabs.harvard.edu/abs/2011CSE....13b..22V}
}

@InProceedings{pandas,
  author    = { Wes McKinney },
  title     = { Data Structures for Statistical Computing in Python },
  booktitle = { Proceedings of the 9th Python in Science Conference },
  pages     = { 51 - 56 },
  year      = { 2010 },
  editor    = { St\'efan van der Walt and Jarrod Millman }
}

@book{Python3, 
 author = {Van Rossum, Guido and Drake, Fred L.}, 
 title = {Python 3 Reference Manual}, 
 year = {2009}, 
 isbn = {1441412697}, 
 publisher = {CreateSpace}, 
 address = {Scotts Valley, CA} 
}

@BOOK{Kippenhahnbook2012,
       author = {{Kippenhahn}, Rudolf and {Weigert}, Alfred and {Weiss}, Achim},
        title = "{Stellar Structure and Evolution}",
         year = 2013,
          doi = {10.1007/978-3-642-30304-3},
       adsurl = {https://ui.adsabs.harvard.edu/abs/2013sse..book.....K}
}

@ARTICLE{Nascimbeni-2022,
       author = {{Nascimbeni}, V. and {Piotto}, G. and {B{\"o}rner}, A. and {Montalto}, M. and {Marrese}, P.~M. and {Cabrera}, J. and {Marinoni}, S. and {Aerts}, C. and {Altavilla}, G. and {Benatti}, S. and {Claudi}, R. and {Deleuil}, M. and {Desidera}, S. and {Fabrizio}, M. and {Gizon}, L. and {Goupil}, M.~J. and {Granata}, V. and {Heras}, A.~M. and {Magrin}, D. and {Malavolta}, L. and {Mas-Hesse}, J.~M. and {Ortolani}, S. and {Pagano}, I. and {Pollacco}, D. and {Prisinzano}, L. and {Ragazzoni}, R. and {Ramsay}, G. and {Rauer}, H. and {Udry}, S.},
        title = "{The PLATO field selection process. I. Identification and content of the long-pointing fields}",
      journal = {\aap},
         year = 2022,
        month = feb,
       volume = {658},
          eid = {A31},
        pages = {A31},
          doi = {10.1051/0004-6361/202142256},
archivePrefix = {arXiv},
       eprint = {2110.13924},
 primaryClass = {astro-ph.EP},
       adsurl = {https://ui.adsabs.harvard.edu/abs/2022A&A...658A..31N}
}

@ARTICLE{Wilson-1941,
       author = {{Wilson}, Raymond Harrison},
        title = "{Construction and use of an interferometer for measurement of close double stars with the eighteen-inch refractor. Continuation of the use of the interferometer for close double star measurements at Flower Observatory}",
      journal = {Publications of the University of Pennsylvania Flower Astronomical Observatory},
         year = 1941,
        month = jan,
       volume = {6},
        pages = {1},
       adsurl = {https://ui.adsabs.harvard.edu/abs/1941PUPFA...6b...1W}
}

@ARTICLE{Labeyrie-1974,
       author = {{Labeyrie}, A. and {Bonneau}, D. and {Stachnik}, R.~V. and {Gezari}, D.~Y.},
        title = "{Speckle Interferometry. III. High-Resolution Measurements of Twelve Close Binary Systems}",
      journal = {\apjl},
         year = 1974,
        month = dec,
       volume = {194},
        pages = {L147},
          doi = {10.1086/181689},
       adsurl = {https://ui.adsabs.harvard.edu/abs/1974ApJ...194L.147L}
}

@ARTICLE{McAlister-1982,
       author = {{McAlister}, H.~A.},
        title = "{Masses and luminosities for the giant spectroscopic speckle interferometric binaries gamma Persei and phi Cygni.}",
      journal = {\aj},
         year = 1982,
        month = mar,
       volume = {87},
        pages = {563-569},
          doi = {10.1086/113130},
       adsurl = {https://ui.adsabs.harvard.edu/abs/1982AJ.....87..563M}
}

@ARTICLE{Bahng-1958,
       author = {{Bahng}, J.~D.~R.},
        title = "{Multicolor photoelectric photometry of stars with composite spectra.}",
      journal = {\apj},
         year = 1958,
        month = nov,
       volume = {128},
        pages = {572},
          doi = {10.1086/146571},
       adsurl = {https://ui.adsabs.harvard.edu/abs/1958ApJ...128..572B}
}

@ARTICLE{Cowley-1976,
       author = {{Cowley}, A.~P.},
        title = "{Spectral classification of the bright F stars.}",
      journal = {\pasp},
         year = 1976,
        month = apr,
       volume = {88},
        pages = {95-110},
          doi = {10.1086/129905},
       adsurl = {https://ui.adsabs.harvard.edu/abs/1976PASP...88...95C}
}

@ARTICLE{Popper-1987,
       author = {{Popper}, Daniel M. and {McAlister}, Harold A.},
        title = "{Gamma Persei--Not Overmassive but Overluminous}",
      journal = {\aj},
         year = 1987,
        month = sep,
       volume = {94},
        pages = {700},
          doi = {10.1086/114507},
       adsurl = {https://ui.adsabs.harvard.edu/abs/1987AJ.....94..700P}
}

@ARTICLE{Grevesse-1984,
       author = {{Grevesse}, N.},
        title = "{Accurate atomic data and solar photospheric spectroscopy.}",
      journal = {Physica Scripta Volume T},
         year = 1984,
        month = jan,
       volume = {8},
        pages = {49-58},
          doi = {10.1088/0031-8949/1984/T8/008},
       adsurl = {https://ui.adsabs.harvard.edu/abs/1984PhST....8...49G}
}

@misc{tarczay_nehez_2026_19470956,
  author       = {Tarczay-Nehéz, Dóra},
  title        = {What's Their Age Again? A Blue-Straggler Merger
                   Scenario in the $\gamma$ Persei Binary System
                  },
  month        = apr,
  year         = 2026,
  publisher    = {Zenodo},
  doi          = {10.5281/zenodo.19470956},
  url          = {https://doi.org/10.5281/zenodo.19470956},
}

@ARTICLE{Berger-2022,
       author = {{Berger}, Travis A. and {van Saders}, Jennifer L. and {Huber}, Daniel and {Gaidos}, Eric and {Schlieder}, Joshua E. and {Claytor}, Zachary R.},
        title = "{Is [Y/Mg] a Reliable Age Diagnostic for FGK Stars?}",
      journal = {\apj},
         year = 2022,
        month = sep,
       volume = {936},
       number = {2},
          eid = {100},
        pages = {100},
          doi = {10.3847/1538-4357/ac8746},
archivePrefix = {arXiv},
       eprint = {2206.10624},
 primaryClass = {astro-ph.SR},
       adsurl = {https://ui.adsabs.harvard.edu/abs/2022ApJ...936..100B}
}

\onecolumn

\begin{appendix}

\section{\texttt{MESA} model grid for the primary component}

\label{app:primar}

\begin{figure}[H]
    \centering
    \includegraphics[width=.98\linewidth]{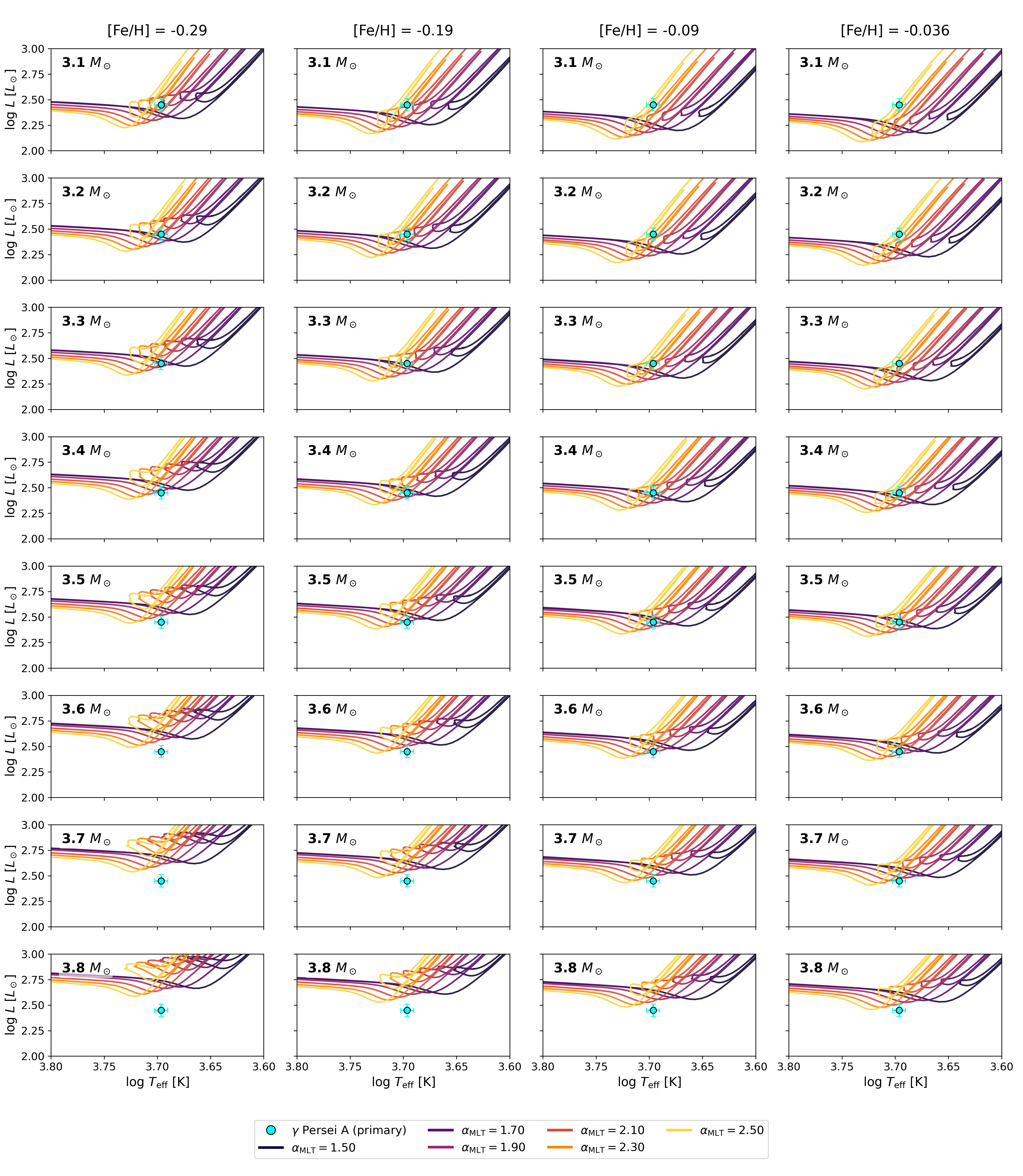}
    \caption{Calculated \texttt{MESA} evolutionary tracks for the primary component of the $\gamma$ Persei system for high overshoot models. Stellar masses increases from top to bottom and metallicites increase from left to right panels. 
    Each evolutionary track is color-coded by the $\alpha_\mathrm{MLT}$ parameter. 
    The cyan circles denote the primary component of the $\gamma$ Persei system derived from observational data.}
    \label{fig:1app1}
\end{figure}

\begin{figure}
    \centering
    \includegraphics[width=\linewidth]{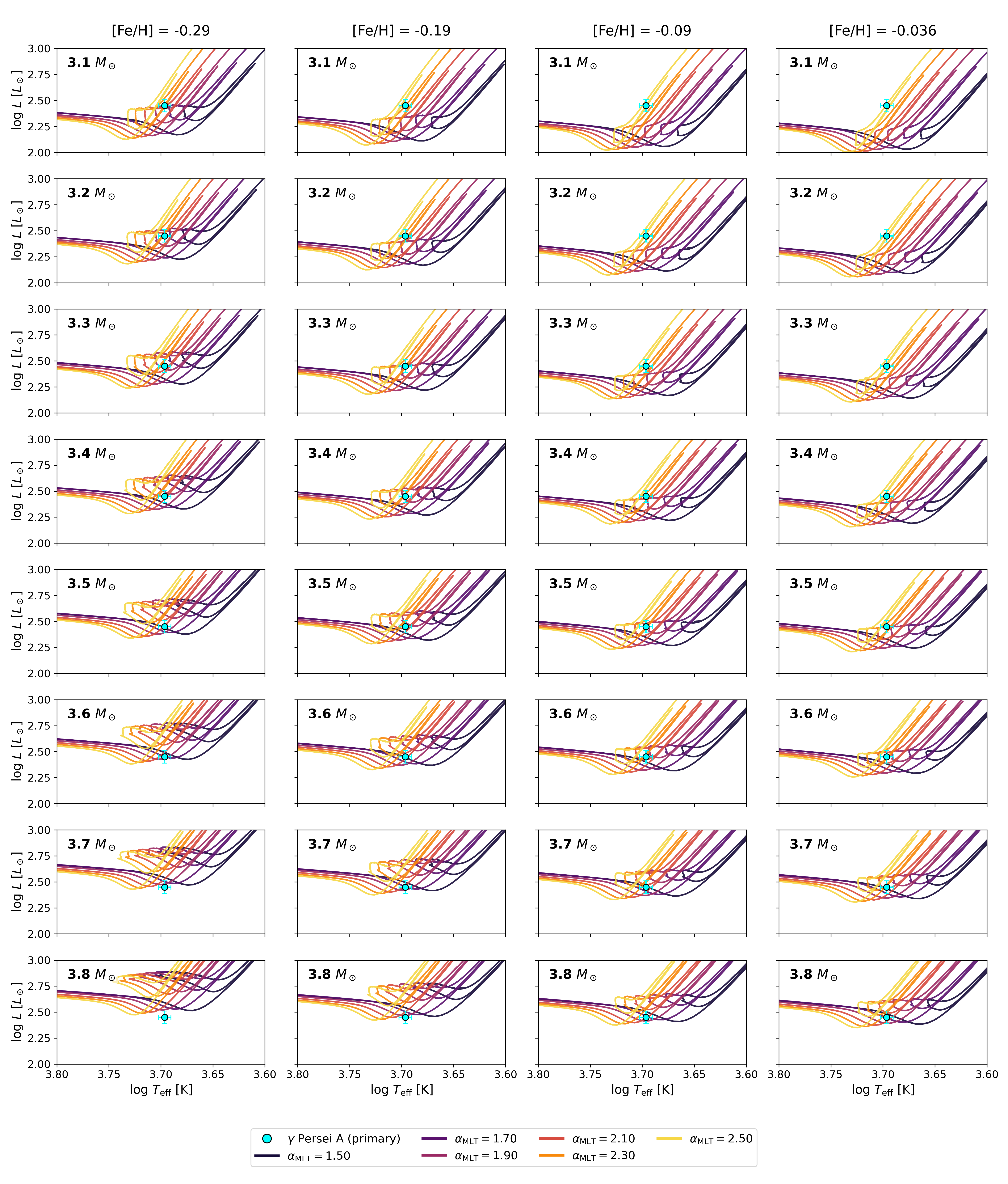}
    \caption{Same as Figure\,\ref{fig:1app1} for moderate overshoot models.}
    \label{fig:1app2}
\end{figure}

\begin{figure}
    \centering
    \includegraphics[width=\linewidth]{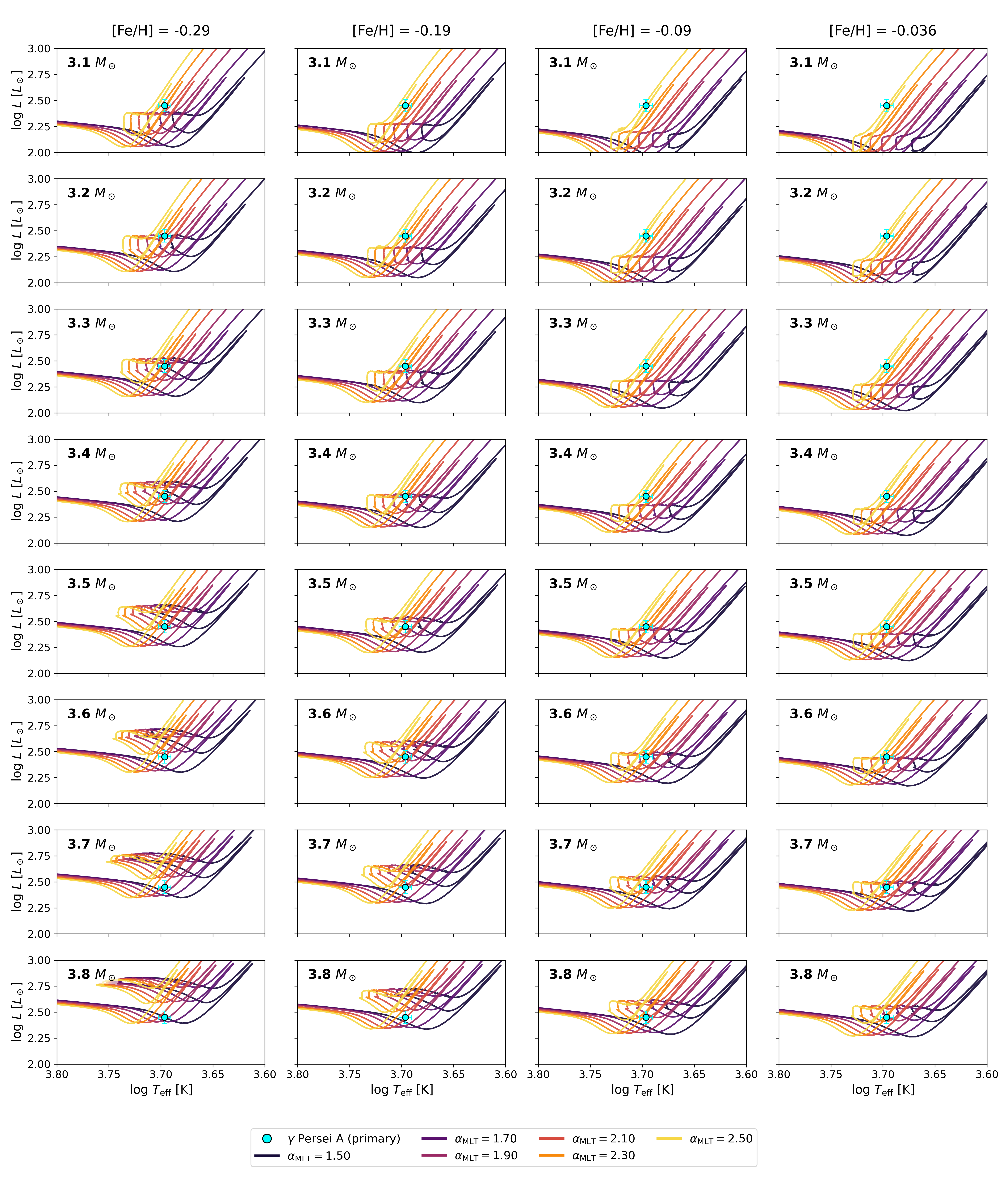}
    \caption{Same as Figure\,\ref{fig:1app1} for low overshoot models.}
    \label{fig:1app3}
\end{figure}

Calculated \texttt{MESA} evolutionary tracks for the primary component of the $\gamma$ Persei system. Figures\,\ref{fig:1app1}, \ref{fig:1app2}, and \ref{fig:1app3} present the high-, moderate-, and low overshoot models with stellar masses increasing from top to bottom and metallicities increasing from left to right panels. 
Each evolutionary track is color-coded by the $\alpha_\mathrm{MLT}$ parameter. 
The cyan circles denote the primary component of the $\gamma$ Persei system derived from observational data.

\FloatBarrier

\section{\texttt{MESA} model grid for the secondary component}

\label{app:secondary}

\begin{figure}[H]
    \centering
    \includegraphics[width=\linewidth]{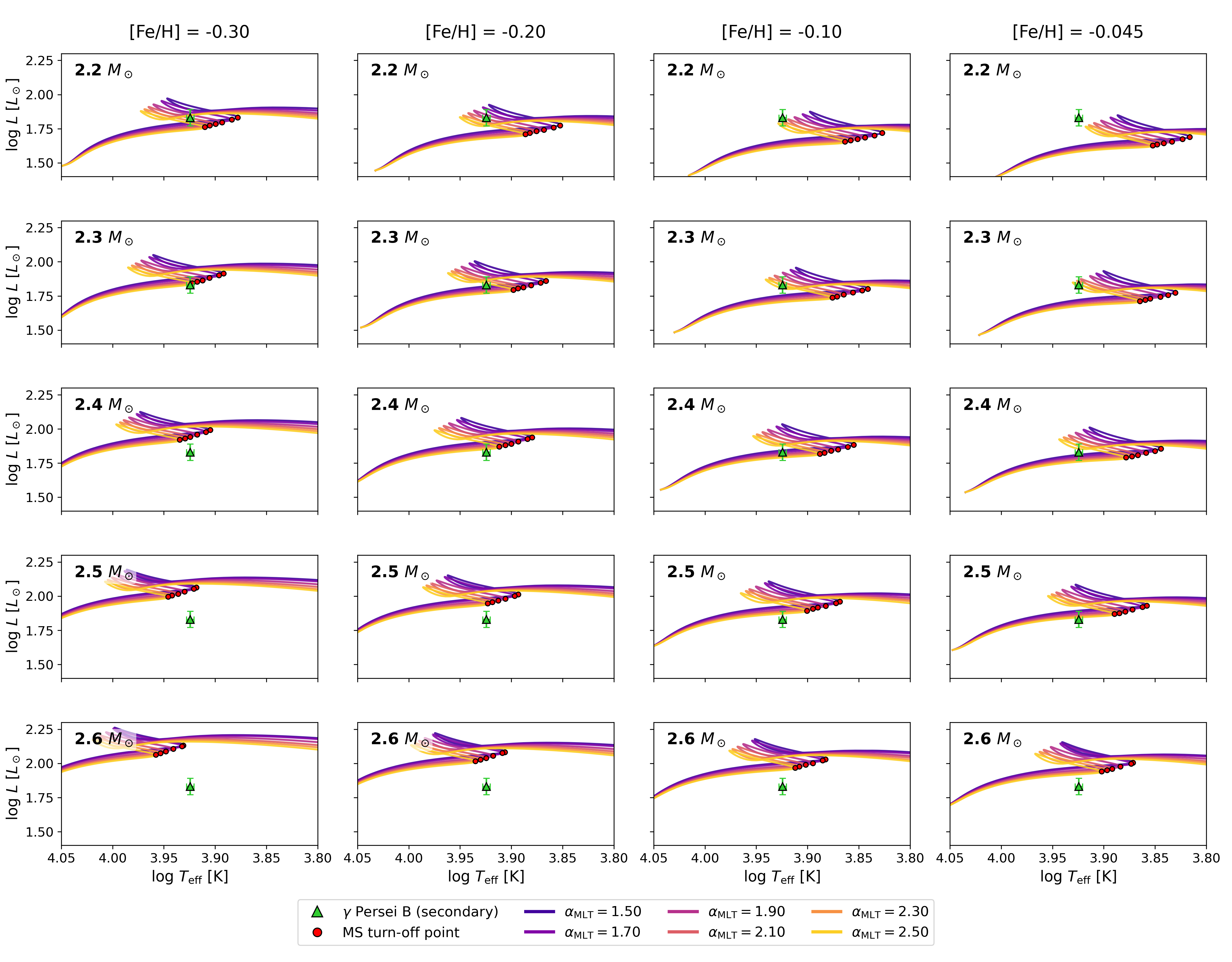}
    \caption{Calculated \texttt{MESA} evolutionary tracks for the secondary component of the $\gamma$ Persei system for high overshoot models. Stellar masses increase from top to bottom and metallicities increase from left to right panels. 
    Each evolutionary track is color-coded by the $\alpha_\mathrm{MLT}$ parameter. 
    The green triangles denote the secondary component of the $\gamma$ Persei system derived from observational data, while the red circle represents the main-sequence turn-off point for each evolutionary track.}
    \label{fig:2app1}
\end{figure}

\begin{figure*}
    \centering
    \includegraphics[width=\linewidth]{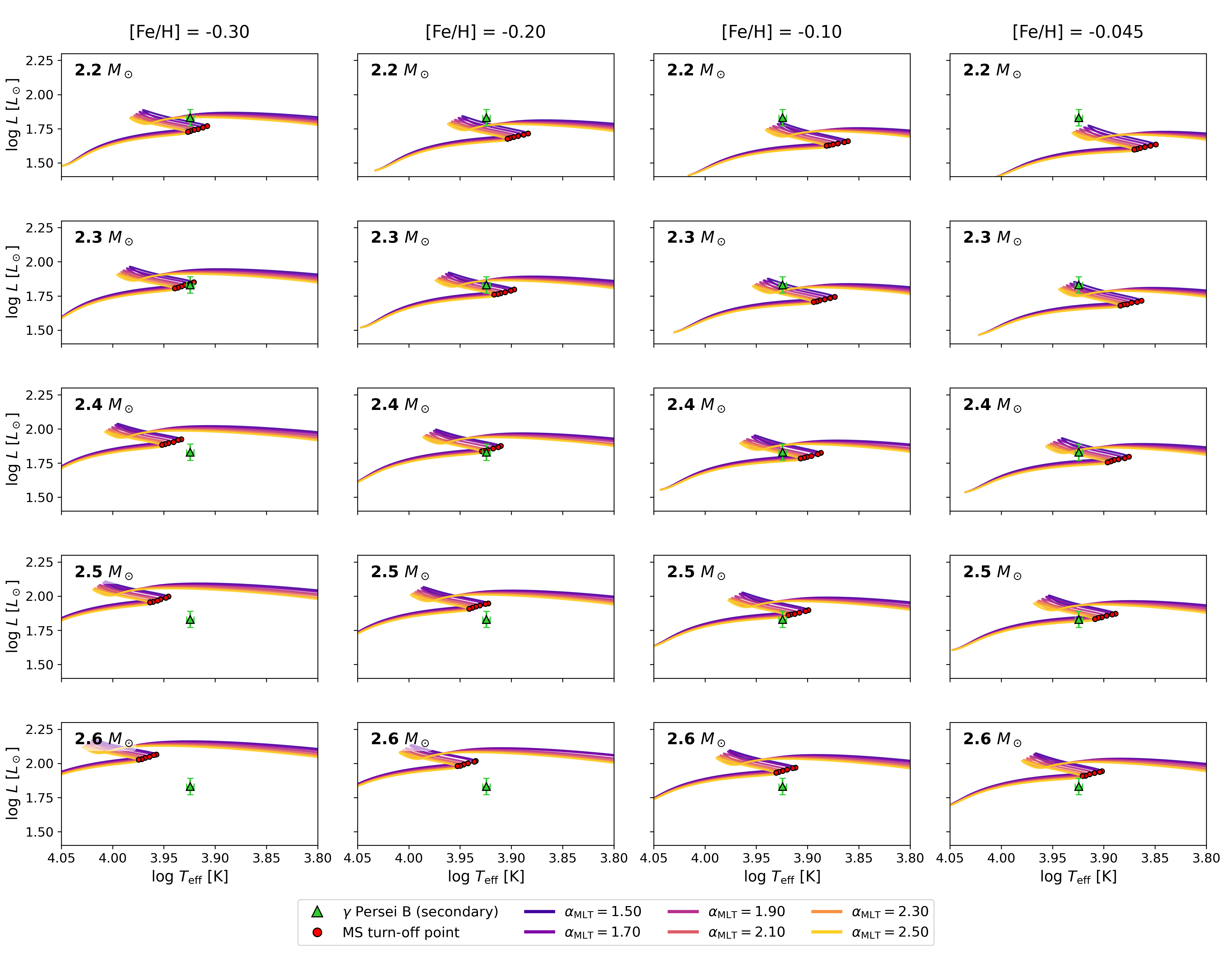}
    \caption{Same as Figure\,\ref{fig:2app1} for moderate overshoot models.}
    \label{fig:2app2}
\end{figure*}

\begin{figure*}
    \centering
    \includegraphics[width=\linewidth]{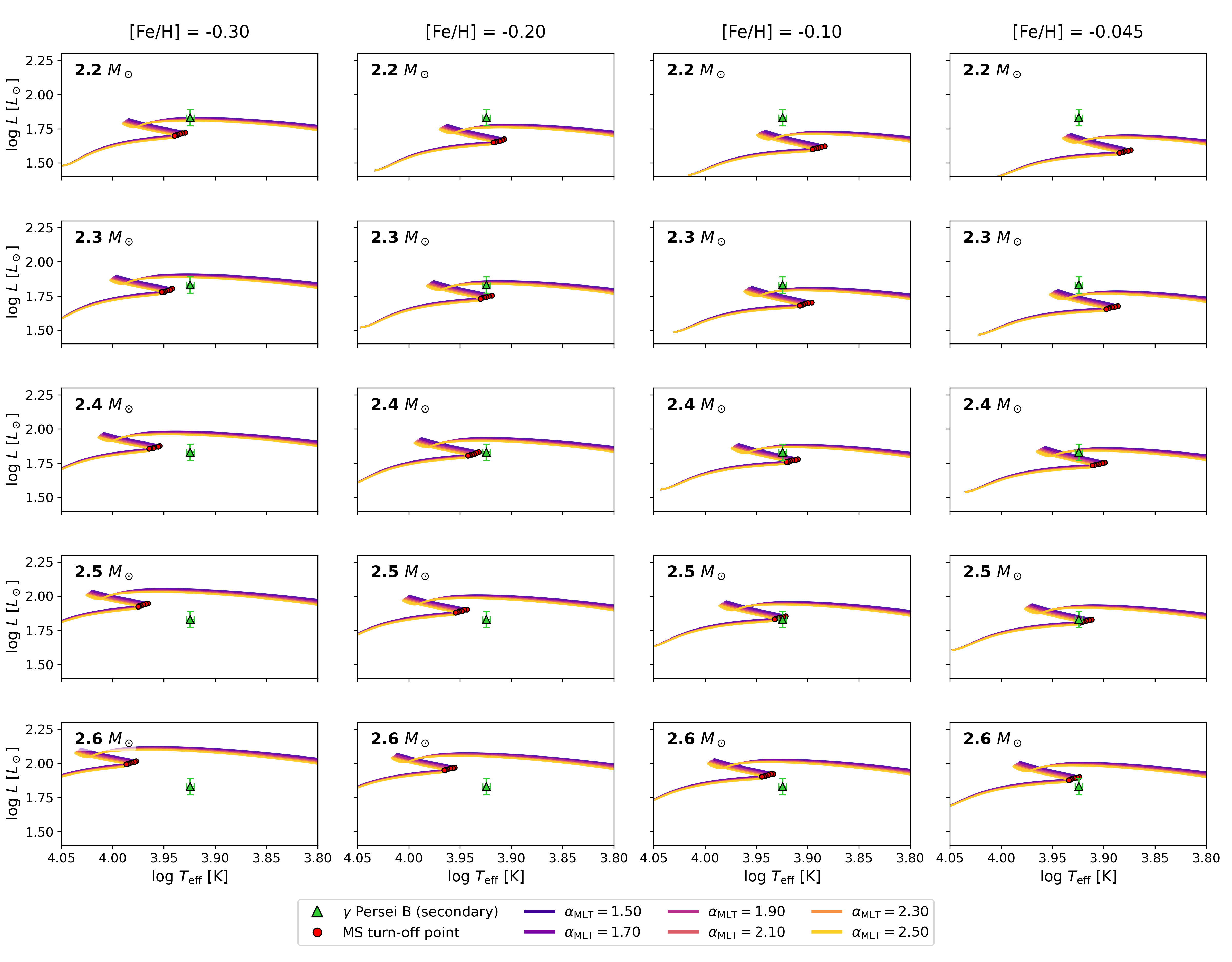}
    \caption{Same as Figure\,\ref{fig:2app1} for low overshoot models.}
    \label{fig:2app3}
\end{figure*}

Calculated \texttt{MESA} evolutionary tracks for the secondary component of the $\gamma$ Persei system. Figures\,\ref{fig:2app1}, \ref{fig:2app2}, and \ref{fig:2app3} present the high-, moderate-, and low overshoot models with stellar masses increasing from top to bottom and metallicities increase from left to right panels. 
Each evolutionary track is color-coded by the $\alpha_\mathrm{MLT}$ parameter. 
The green triangles denote the secondary component of the $\gamma$ Persei system derived from observational data, while the red circle represents the main-sequence turn-off point for each evolutionary track.

\FloatBarrier

\renewcommand{\thetable}{C.1}

\section{MESA model constraints from the evolutionary grid}

\label{app:table}

\vskip 2.6em

\tablefirsthead{
    \toprule \toprule \multicolumn{4}{c|}{\textit{Primary component}} & \multicolumn{4}{c}{\textit{Secondary component}}\\
    
    [Fe/H]$_1$ & $t_1$ [Myr] & $M_1$ [$M_\odot$] & $R_1$ [$R_\odot$] & [Fe/H]$_2$ & $t_2$ [Myr]  & $M_2$ [$M_\odot$] & $R_2$ [$R_\odot$] \\ \midrule
        }
    
\tablecaption{Age and mass ranges where the two components of the system intersect MESA evolutionary tracks (within the observational errors). Ages are given in megayears. \label{tab:mesa_results}}

  \tablehead{
    \textbf{Table C.1.} continued. \\ \hline\hline
    \multicolumn{4}{c|}{\textit{Primary component}} & \multicolumn{4}{c}{\textit{Secondary component}}\\
    
    [Fe/H]$_1$ & $t_1$ [Myr] & $M_1$ [$M_\odot$] & $R_1$ [$R_\odot$] & [Fe/H]$_2$ & $t_2$ [Myr]  & $M_2$ [$M_\odot$] & $R_2$ [$R_\odot$] \\ \hline
  }

\renewcommand{\arraystretch}{1.2}
\begin{center}
\small
\begin{supertabular}{cccc|cccc}
    \multicolumn{8}{c}{\textbf{High overshoot models}} \\ \hline
    
    \multicolumn{8}{c}{\textit{$\alpha_{\rm MLT} = 1.5$}} \\
    $-0.29$ & 295--316 & 3.2--3.3 & 22.041 -- 23.661 & -0.30 & 708--776 & 2.2--2.6 & 3.732 -- 3.789 \\
    $-0.19$ & 288--316 & 3.3--3.4 & 22.006 -- 23.739 & -0.20 & 724 & 2.3 & 3.840\\
    $-0.09$ & 282--309 & 3.4--3.5 & 22.270 -- 24.007 & -0.10 & 691 & 2.4 & 3.949  \\
    $-0.036$ & 294--317 & 3.4--3.6 & 21.727 -- 23.412 &  -0.045 &  694 & 2.4 &  3.795\\ \hline

    \multicolumn{8}{c}{\textit{$\alpha_{\rm MLT} = 1.7$}} \\
    $-0.29$ & 275--316 & 3.2--3.4 & 21.323 -- 22.778 & -0.30 & 691--831 & 2.2--2.6 & 3.713 -- 4.064 \\
    $-0.19$ & 269--316 & 3.3--3.5 & 21.253 -- 22.929 & -0.20 & 708--891 & 2.2--2.3 & 3.760 -- 3.850 \\
    $-0.09$ & 288--309 & 3.4--3.5 & 21.504 -- 23.130 & -0.10 & 676 & 2.4 & 3.837 -- 3.928  \\
    $-0.036$ &  273--294 & 3.4--3.6 & 22.372 -- 23.943 &  -0.045 &  683 & 2.4--2.5 &  3.767\\ \hline

    \multicolumn{8}{c}{\textit{$\alpha_{\rm MLT} = 1.9$}} \\
    $-0.29$ & 269--288 & 3.3--3.4 & 21.695 -- 23.422 & -0.30 & 676--794 & 2.2--2.3 & 3.920 -- 4.049 \\
    $-0.19$ & 245--282 & 3.4--3.6 & 21.227 -- 23.677 & -0.20 & 645--871 & 2.2--2.6 & 3.708 -- 4.095 \\
    $-0.09$ & 245--282 & 3.5--3.7 & 21.280 -- 22.982 & -0.10 & 660 & 2.4 & 3.754 -- 3.838 \\
    $-0.036$ &  250--269 & 3.5--3.7 & 22.136 -- 23.995 &  -0.045 &  618 & 2.4--2.5 &  4.062\\ \hline

    \multicolumn{8}{c}{\textit{$\alpha_{\rm MLT} = 2.1$}} \\
    $-0.29$ & 263--316 & 3.2--3.4 & 21.646 -- 23.990 & -0.30 & 660--794 & 2.2--2.3 & 3.854 -- 4.001 \\
    $-0.19$ & 239--363 & 3.2--3.6 & 21.369 -- 24.189 & -0.20 & 631--851 & 2.2--2.6 & 3.722 -- 4.092 \\
    $-0.09$ & 239--389 & 3.2--3.7 & 21.209 -- 24.163 & -0.10 & 602--812 & 2.3--2.5 & 3.789 -- 4.095 \\
    $-0.036$ &  229--402 & 3.2--3.8 & 21.227 -- 24.195 &  -0.045 &  606--841 & 2.3--2.5 &  3.703 --  4.100\\ \hline

    \multicolumn{8}{c}{\textit{$\alpha_{\rm MLT} = 2.3$}} \\
    $-0.29$ & 282--309 & 3.2--3.3 & 22.093 -- 24.100 & -0.30 & 660--776 & 2.2--2.3 & 3.759 -- 4.094 \\
    $-0.19$ & 257--363 & 3.2--3.5 & 21.287 -- 24.179 & -0.20 & 616--831 & 2.2--2.4 & 3.718 -- 4.067 \\
    $-0.09$ & 235--389 & 3.2--3.7 & 21.231 -- 24.089 & -0.10 & 588--891 & 2.2--2.5 & 3.720 -- 4.093 \\
    $-0.036$ &  241--439 & 3.1--3.8 & 21.239 -- 24.147 &  -0.045 &  600--828 & 2.3--2.5 &  3.941 --  4.079\\ \hline

    \multicolumn{8}{c}{\textit{$\alpha_{\rm MLT} = 2.5$}} \\
    -- & -- & -- & -- & -0.30 & 660--758 & 2.2--2.6 & 3.727 -- 4.007 \\
    $-0.19$ & 316--316 & 3.2--3.2 & 21.329 -- 24.149 & -0.20 & 616--812 & 2.2--2.4 & 3.908 -- 3.996 \\
    $-0.09$ & 263--389 & 3.2--3.5 & 21.329 -- 24.149 & -0.10 & 588--871 & 2.2--2.5 & 3.715 -- 4.064 \\
    $-0.036$ &  274--439 & 3.1--3.6 & 21.250 -- 24.197 &  -0.045 &  593--814 & 2.3--2.5 &  3.811 --  4.083\\ \hline

    \multicolumn{8}{c}{\textbf{Med overshoot models}} \\ \hline
    \multicolumn{8}{c}{\textit{$\alpha_{\rm MLT} = 1.5$}} \\
    $-0.29$ & 219--235 & 3.5--3.6 & 22.324 -- 23.771 & -0.30 & 660--758 & 2.2--2.6 & 3.732 -- 4.034 \\
    $-0.19$ & 213--245 & 3.5--3.7 & 22.438 -- 23.880 & -0.20 & 616--831 & 2.2--2.4 & 3.720 -- 4.026 \\
    $-0.09$ & 213--245 & 3.6--3.8 & 21.403 -- 24.080 & -0.10 & 588--891 & 2.2--2.5 & 3.700 -- 4.091 \\
    $-0.036$ &  219--234 & 3.6--3.8 & 22.198 -- 23.539 &  -0.045 &  595--820 & 2.3--2.5 &  3.855 --  4.064\\ \hline

    \multicolumn{8}{c}{\textit{$\alpha_{\rm MLT} = 1.7$}} \\
    $-0.29$ & 204--323 & 3.2--3.7 & 21.349 -- 24.158 & -0.30 & 645--758 & 2.2--2.6 & 3.707 -- 4.029 \\
    $-0.19$ & 204--229 & 3.6--3.8 & 21.370 -- 24.115 & -0.20 & 616--812 & 2.2--2.4 & 3.887 -- 4.012 \\
    $-0.09$ & 213--229 & 3.7--3.8 & 21.592 -- 22.994 & -0.10 & 575--871 & 2.2--2.5 & 3.716 -- 4.081 \\
    $-0.036$ &  219--219 & 3.7--3.8 & 22.463 -- 22.506 &  -0.045 &  589--806 & 2.3--2.5 &  3.739 --  4.092\\ \hline

    \multicolumn{8}{c}{\textit{$\alpha_{\rm MLT} = 1.9$}} \\
    $-0.29$ & 204--331 & 3.2--3.7 & 21.423 -- 24.181 & -0.30 & 741 & 2.2 & 3.962 -- 3.962 \\
    $-0.19$ & 200--355 & 3.2--3.8 & 21.208 -- 23.908 & -0.20 & 602--794 & 2.2--2.6 & 3.777 -- 4.092 \\
    $-0.09$ & 209--316 & 3.4--3.8 & 21.309 -- 24.199 & -0.10 & 575--758 & 2.3--2.5 & 3.700 -- 4.063 \\
    $-0.036$ &  216--326 & 3.4--3.8 & 21.519 -- 24.184 &  -0.045 &  583--791 & 2.3--2.5 &  3.737 --  3.998\\ \hline

    \multicolumn{8}{c}{\textit{$\alpha_{\rm MLT} = 2.1$}} \\
    $-0.29$ & 200--331 & 3.2--3.7 & 21.226 -- 24.129 & -0.30 & 724 & 2.2 & 3.939 -- 3.939 \\
    $-0.19$ & 194--355 & 3.2--3.8 & 21.275 -- 24.134 & -0.20 & 602--776 & 2.2--2.6 & 3.706 -- 4.064 \\
    $-0.09$ & 209--355 & 3.3--3.8 & 21.348 -- 24.126 & -0.10 & 562--741 & 2.3--2.5 & 3.708 -- 4.079 \\
    $-0.036$ &  213--334 & 3.4--3.8 & 21.221 -- 24.180 &  -0.045 &  571--695 & 2.3--2.5 &  3.792 --  3.991\\ \hline

    \multicolumn{8}{c}{\textit{$\alpha_{\rm MLT} = 2.3$}} \\
    $-0.29$ & 224--331 & 3.2--3.5 & 21.428 -- 23.985 & -0.30 & 708 & 2.2 & 3.876 -- 3.967 \\
    $-0.19$ & 209--355 & 3.2--3.7 & 21.214 -- 24.143 & -0.20 & 588--776 & 2.2--2.4 & 3.702 -- 4.094 \\
    $-0.09$ & 204--380 & 3.2--3.8 & 21.264 -- 24.191 & -0.10 & 562--741 & 2.3--2.5 & 3.715 -- 4.088 \\
    $-0.036$ &  210--435 & 3.1--3.8 & 21.260 -- 24.166 &  -0.045 &  570--685 & 2.3--2.5 &  3.817 --  4.088\\ \hline

    \multicolumn{8}{c}{\textit{$\alpha_{\rm MLT} = 2.5$}} \\
    -- & -- & -- & -- & -0.30 & 708 & 2.2--2.6 & 3.900 -- 3.900 \\
    $-0.19$ & 269--295 & 3.2--3.3 & 21.392 -- 24.165 & -0.20 & 588--758 & 2.2--2.4 & 3.711 -- 4.014 \\
    $-0.09$ & 251--380 & 3.2--3.5 & 21.392 -- 24.165 & -0.10 & 550--724 & 2.3--2.5 & 3.704 -- 4.097 \\
    $-0.036$ &  222--437 & 3.1--3.7 & 21.224 -- 24.185 &  -0.045 &  563--675 & 2.4--2.6 &  3.777 --  4.068\\ \hline

    \multicolumn{8}{c}{\textbf{Low overshoot models}} \\ \hline
    \multicolumn{8}{c}{\textit{$\alpha_{\rm MLT} = 1.5$}} \\
    $-0.29$ & 178--309 & 3.2--3.8 & 21.448 -- 24.178 & -0.30 & 708 & 2.2--2.6 & 3.732 -- 3.935 \\
    $-0.19$ & 186 & 3.8 & 21.455 -- 21.535 & -0.20 & 676--758 & 2.2--2.3 & 3.954 -- 4.028 \\
    -- & -- & -- & -- & -0.10 & 550--724 & 2.3--2.5 & 3.715 -- 4.052 \\
    -- & -- & -- & -- &  -0.045 &  557--670 & 2.4--2.6 &  3.732 --  4.010\\ \hline

    \multicolumn{8}{c}{\textit{$\alpha_{\rm MLT} = 1.7$}} \\
    $-0.29$ & 178--323 & 3.2--3.8 & 21.213 -- 23.175 & -0.30 & 691 & 2.2--2.6 & 3.727 -- 3.818 \\
    $-0.19$ & 229--316 & 3.3--3.6 & 21.258 -- 24.181 & -0.20 & 660--741 & 2.2--2.3 & 3.931 -- 4.003 \\
    $-0.09$ & 224 & 3.7 & 24.063 -- 24.063 & -0.10 & 550--708 & 2.3--2.5 & 3.708 -- 4.065 \\
    -- & -- & -- & -- &  -0.045 &  517--662 & 2.4--2.6 &  3.703 --  4.089\\ \hline

    \multicolumn{8}{c}{\textit{$\alpha_{\rm MLT} = 1.9$}} \\
    $-0.29$ & 178--331 & 3.2--3.8 & 21.353 -- 24.055 & -0.30 & 691 & 2.2 & 3.824 -- 3.824 \\
    $-0.19$ & 186--323 & 3.3--3.8 & 21.421 -- 23.872 & -0.20 & 660 & 2.3--2.6 & 3.969 -- 3.969 \\
    $-0.09$ & 209--288 & 3.5--3.8 & 21.203 -- 23.877 & -0.10 & 550--708 & 2.3--2.5 & 3.709 -- 4.084 \\
    $-0.036$ &  213--276 & 3.6--3.8 & 21.251 -- 24.193 &  -0.045 &  515--652 & 2.4--2.6 &  3.710 --  4.092\\ \hline

    \multicolumn{8}{c}{\textit{$\alpha_{\rm MLT} = 2.1$}} \\
    $-0.29$ & 175--331 & 3.2--3.8 & 21.238 -- 24.195 & -- & -- & -- & -- \\
    $-0.19$ & 182--323 & 3.3--3.8 & 21.234 -- 24.171 & -0.20 & 645 & 2.3--2.6 & 3.914 -- 4.005 \\
    $-0.09$ & 195--316 & 3.4--3.8 & 21.201 -- 24.147 & -0.10 & 616--691 & 2.3--2.4 & 3.734 -- 4.076 \\
    $-0.036$ &  200--303 & 3.5--3.8 & 21.241 -- 24.064 &  -0.045 &  547--646 & 2.4--2.5 &  3.711 --  3.981\\ \hline

    \multicolumn{8}{c}{\textit{$\alpha_{\rm MLT} = 2.3$}} \\
    $-0.29$ & 200--331 & 3.2--3.6 & 21.308 -- 24.113 & -0.30 & 676 & 2.2 & 3.839 -- 3.839 \\
    $-0.19$ & 182--355 & 3.2--3.8 & 21.237 -- 24.196 & -0.20 & 645 & 2.3 & 3.945 -- 3.945 \\
    $-0.09$ & 190--380 & 3.2--3.8 & 21.201 -- 24.135 & -0.10 & 602--691 & 2.3--2.4 & 3.703 -- 4.091 \\
    $-0.036$ &  198--434 & 3.1--3.8 & 21.241 -- 24.137 &  -0.045 &  546--639 & 2.4--2.5 &  3.712 --  3.966\\ \hline

    \multicolumn{8}{c}{\textit{$\alpha_{\rm MLT} = 2.5$}} \\
    $-0.29$ & 263--269 & 3.2 & 22.848 -- 23.551 & -0.30 & 602--676 & 2.2--2.6 & 3.791 -- 3.791 \\
    $-0.20$ & 224--355 & 3.2--3.5 & 21.447 -- 24.115 & -0.20 & 631 & 2.3 & 3.860 -- 3.977 \\
    $-0.10$ & 219--380 & 3.2--3.6 & 21.245 -- 24.200 & -0.10 & 602--691 & 2.3--2.4 & 3.701 -- 4.073 \\
    $-0.036$ &  197--434 & 3.1--3.8 & 21.208 -- 24.170 &  -0.045 &  544--634 & 2.4--2.5 &  3.713 --  3.978\\ \hline
\end{supertabular}
\end{center}

\end{appendix}

\end{document}